\documentclass[11pt]{article}
\usepackage{amsthm}
\usepackage{latexsym}
\usepackage{amssymb}
\usepackage{amsmath}
\usepackage{mathrsfs}
\usepackage{array}
\usepackage{euscript}
\usepackage[T1]{fontenc}
\newcommand{\Csp}{\hspace{0.09in}} 
\newcommand{\CPsp}{\hspace{0.135in}} 

\newcommand{\ba}{\mathbf{a}}
\newcommand{\bb}{\mathbf{b}}
\newcommand{\bc}{\mathbf{c}}
\newcommand{\bd}{\mathbf{d}}

\newcommand{\R}{\mathbb{R}}
\newcommand{\N}{\mathbb{N}}

\newcommand{\p}{\partial}
\renewcommand{\d}{\mathrm{d}}

\newcommand{\dvol}{\mathrm{dVol}}
\newcommand{\hook}{\lrcorner}
\newcommand{\hnabla}{\hat{\nabla}}

\def\hpsi{\hat{\psi}}

\newcommand{\thorn}{\mbox{\th}}

\newcommand{\scri}{{\mathscr I}}

\newtheorem{theorem}{Theorem}
\newtheorem{proposition}{Proposition}[section]

\newtheorem{lemma}{Lemma}[section]
\newtheorem{remark}{Remark}[section]
\begin{document}\mbox{}

\vspace{0.25in}

\begin{center}
{\huge{Peeling of Dirac and Maxwell fields on a Schwarzschild background}}

\vspace{0.25in}

{\large{Lionel J. MASON\footnote{The Mathematical Institute, 24-29 St
    Giles', OXFORD OX1 3LB, UNITED KINGDOM. \\ \noindent
    lmason@maths.ox.ac.uk} \&
Jean-Philippe NICOLAS\footnote{Laboratoire de Math\'ematiques, Universit\'e de Brest, 6 avenue Victor Le Gorgeu, 29200 BREST,
    FRANCE. \\ \noindent Jean-Philippe.Nicolas@univ-brest.fr}} }
\end{center}

\begin{abstract}
We study the peeling of Dirac and Maxwell fields on a Schwarzschild background following the approach developed by the authors in \cite{MaNi2009} for the wave equation. The method combines a conformal compactification with vector field techniques in order to work out the optimal space of initial data for a given transverse regularity of the rescaled field across null infinity. The results show that analogous decay and regularity assumptions in Minkowski and in Schwarzschild produce the same regularity across null infinity. The results are valid also for the classes of asymptotically simple spacetimes constructed by Corvino-Schoen / Chrusciel-Delay.
\end{abstract}

\tableofcontents

\section{Introduction}

Zero rest-mass fields on asymptotically flat spacetimes admit a peeling-off, or peeling, property, discovered by Sachs in 1961 \cite{Sa61} in the flat case. It can be described in terms of principle null directions~: for a zero rest-mass field $\phi_{AB...F} = \phi_{(AB...F)}$ with $n$ indices, the part of the field falling-off like $r^{-k-1}$, $0\leq k \leq n$, along outgoing null geodesics, has $n-k$ of its principle null directions aligned along the generator of the geodesics. On Minkowski spacetime, this can be stated simply in terms of components in a well-chosen spin-frame. Consider the Newman-Penrose tetrad
\[ l = \frac{1}{\sqrt{2}} (\partial_t + \partial_r) \, ,~ n = \frac{1}{\sqrt{2}} (\partial_t - \partial_r) \, ,~ m = \frac{1}{r\sqrt{2}} (\partial_\theta + \frac{i}{\sin \theta} \partial_\varphi ) \, ,~\bar{m} \]
and the associated spin-frame $\{ o \, ,~ \iota \}$ (unique modulo overall sign), then the component $\phi_{n-k}$, which is the contraction of $\phi$ with $n-k$ $\iota$'s and $k$ $o$'s, falls off like $r^{-k-1}$ along the integral curves of $l$.

It is essential to note that the property as we have brutally stated it is wrong. One can consider initial data on a spacelike slice that are exponentially increasing at infinity and the associated solution will not fall-off at all along outgoing null geodesics. The peeling is true in flat space-time for smooth compactly supported initial data and for certain classes of data that satisfy adequate regularity and fall-off assumptions. The situation is expected to be similar on generic asymptotically flat spacetimes but it has been speculated that the conditions on the initial data may be more stringent due to the more complicated asymptotic structure.

In 1965, Penrose \cite{Pe65} presented a new derivation of the peeling based on conformal compactifications. Using the conformal embedding of Minkowski spacetime into the Einstein cylinder and the conformal invariance of zero rest-mass field equations, he showed that the peeling property is equivalent to the continuity of the rescaled field at null infinity ($\scri$). Then he argued that the peeling should be a generic behaviour of zero rest-mass fields on asymptotically flat spacetimes. He went on to define a class of spacetimes, referred to as asymptotically simple spacetimes, providing a generic model of asymptotic flatness as well as a framework in which the peeling should occur under reasonable conditions. These spacetimes are classifed according to the regularity of their conformal metric at null infinity, which encodes the information of the peeling of the Weyl tensor.

His results raised criticisms regarding the genericity of
asymptotically simple spacetimes and the peeling property. The jist of
the arguments put forward was that the asymptotic structure of the
Schwarzschild metric would only allow peeling for a considerably smaller
family of initial data than in Minkowski spacetime. Indeed,
asymptotically simple spacetimes are meant to be physically reasonable
and as such contain mass-energy, which means that the physical metric
differs from the flat one at first approximation by a
Schwarzschild-type behaviour in $m/r$. The asymptotic structure of
Schwarzschild's spacetime is substantially different
from Minkowski's in particular with a singular conformal structure at
spatial infinity.  It was unclear as to whether this would impose more
stringent hypotheses on initial data to ensure peeling; the
singularity at spatial infinity could interact with the tail of the
falloff of the initial data at spatial infinity and prevent the
peeling that might otherwise take place if the data were compactly
suppported.   The genericity
of the peeling for zero rest-mass fields being questioned, the peeling
for linearized gravity on Schwarzschild's spacetime was also doubted
and this made the asymptotically simple spacetime model appear as
anything but generic. The question of regularity of null infinity and
asymptotic simplicity has now been resolved in various ways,
Christodoulou-Klainerman \cite{ChriKla}, Corvino \cite{Co2000},
Chrusciel and Delay \cite{ChruDe2002, ChruDe2003}, Corvino-Schoen
\cite{CoScho2003}, Friedrich (see \cite{HFri2004} for a survey of his
contributions) and Klainerman-Nicol\`o
\cite{KlaNi,KlaNi2002,KlaNi2003}. However, even in the simple case of
the Schwarzschild metric, it was not at all clear, until the authors
provided a first element of answer in \cite{MaNi2009}, whether zero
rest-mass fields admit peeling properties for reasonably large classes
of initial data. 

Penrose's constructions in \cite{Pe65} provide powerful techniques to analyze the constraints on the initial data implicit in the peeling property. A $t=$ {\it constant} slice of Minkowski spacetime corresponds to a $3$-sphere with a point removed on the Einstein cylinder. Physical initial data that once rescaled extend as smooth functions on the whole $3$-sphere give rise, as ensured by Leray's theorem, to a rescaled solution that is smooth on the whole Einstein cylinder and consequently the physical solution satisfies the peeling property. This is a little crude since the peeling really means continuity of the rescaled solution at null infinity and the class of data considered above provides solutions that are ${\cal C}^\infty$ across $\scri$. A more detailed understanding requires to consider intermediate regularities, but ${\cal C}^k$ spaces are not adapted to the Cauchy problem for hyperbolic equations~; a typical example is the wave equation on Minkowski spacetime~: for initial data $f\vert_{t=0} \in {\cal C}^k (\R^3 )$ and $\partial_t f \vert_{t=0} \in {\cal C}^{k-1} (\R^3 )$, the solution $f$ is generally not in ${\cal C}^k (\R \times \R^3 )$. It is better to use Sobolev spaces instead since they are naturally controlled by energy estimates. In a previous paper \cite{MaNi2009}, the authors used a combination of Penrose's conformal techniques and geometric energy estimates (or vector field methods) to characterize completely the spaces of initial data ensuring peeling for the scalar wave equation on the Schwarzschild metric. A new definition of peeling with specifyable order of regularity was given in terms of (weighted) Sobolev spaces. The classes of data were shown to have analogous fall-off properties as those obtained in the flat case using the conformal embedding in the Einstein cylinder. This established that at least for the wave equation, the different asymptotic structure of the Schwarzschild spacetime does not change the classes of data ensuring peeling. To gain a more complete understanding of the question, it is necessary to study other types of fields on a Schwarzschild background, such as higher spin zero rest-mass fields, or solutions of non linear equations, and then to extend the results to more general asymptotically flat spacetimes.

The present paper uses similar techniques to investigate the peeling for Dirac and Maxwell fields on the Schwarzschild spacetime. It is organized as follows. Section \ref{GeoSet} describes the geometric ingredients of the method~: the conformal (partial) compactification of Schwarzschild's spacetime with choices of Newman-Penrose tetrads on the physical and rescaled spacetimes, the corresponding rescaling of  spin-coefficients, the neighbourhood of spacelike infinity in which we establish our estimates, a choice of foliation and the Morawetz vector field. This vector field is as crucial for the Maxwell case as it was for the wave equation, but it does not play any part in the construction for Dirac fields thanks to the existence of a conserved current independent of a choice of observer. In section \ref{DirMaxFields}, we focus on the field equations, their conformal invariance, the proof of the equivalence between the peeling and the continuity of the rescaled field at null infinity as Penrose gave it in \cite{Pe65}, the conserved quantities (current for Dirac and stress-energy tensor for Maxwell), the associated energies on the hypersurfaces we work with and the generic method for obtaining energy estimates for a perturbed equation (explained simply in the Dirac case). Section \ref{Peeling} contains the peeling results and their proof for Dirac and Maxwell fields. We establish energy estimates for fields supported away from spacelike infinity and then use these to construct our spaces of initial data by completion in the norms obtained on the initial hypersurface. We get a full set of function spaces with all degrees of regularity~; for data in these spaces, we have estimates both ways between the norm on the initial hypersurface and a corresponding norm on $\scri^+$ involving transverse derivatives as well as angular ones. The main difficulty is of course the estimates on transverse derivatives. The control of angular derivatives is straightforward thanks to the spherical symmetry~; in the case of Dirac fields, it can be obtained elegantly by commuting into the equation the Dirac operator on the sphere~; for the Maxwell system, this cannot be done because we have four equations and three unknowns (this is also manifest in the spin and boost weights of the components of the field and their transformation under the Geroch-Held-Penrose angular operators), so we use a set of three Killing vectors on the sphere instead. The main results, given in theorems \ref{DiracPeeling} and \ref{PeelingMaxwell}, are the energy estimates and the function spaces on the initial data surface that one can infer from them. The interpretation of the results, namely the comparison between our classes of initial data and the classes obtained in Minkowski spacetime using the full conformal embedding in the Einstein cylinder, is given in section \ref{Interpretation}, together with a remark on the constraints. All the calculations in this paper are done using partial derivatives of the field components. One may prefer using covariant derivatives of the fields and then taking components. This requires to know the different curvature spinors of the rescaled Schwarzschild spacetime. The calculation of these quantities and the derivation of an energy estimate for a transverse derivative using this approach is given for Dirac fields in the appendix. Although the calculations are more involved, the error terms in the conservation law for the transverse derivative are much simpler than if we use partial derivatives (see remark \ref{PurelyTransverse} at the end of the paper).

Since the main nontrivial results of this paper concern the behaviour of fields in a neighbourhood of space-like infinity and its intersection with null infiinity, the results are valid for the classes of spacetimes of Corvino-Schoen / Chrusciel-Delay.

{\bf Notations.} Throughout the paper, we use the formalisms of abstract indices, $2$-component spinors, Newman-Penrose and Geroch-Held-Penrose.

\section{Geometric setting} \label{GeoSet}

\subsection{Rescaled Schwarzschild spacetime}

We work on the Schwarzschild metric
\begin{gather*}
g = F(r) \d t^2 - F(r)^{-1} \d r^2- r^2 \d \omega^2 \, , ~m>0 \, ,\\
F(r) = 1-2m/r \, ,~\d \omega^2 = \d \theta^2 + \sin^2 \theta \d \varphi^2 \, ,
\end{gather*}
in the region outside the black-hole $\R_t \times ]2m ,+\infty [_r \times S^2_\omega$. Introducing the variables $u=t-r_*$, where $r_*= r +2m \log (r-2m)$ is the Regge-Wheeler coordinate, and $R = 1/r$, we obtain the following expression for the metric $g$ conformally rescaled using the conformal factor $R$~:
\begin{equation} \label{RescMet}
\hat{g}= R^2 g= R^2F\d u^2-2\d u\d R-\d\omega^2 \, ,~ \mbox{still denoting } F = F(r) = 1-2mR \, .
\end{equation}
These choices of conformal rescaling and variables allow to define naturally future null infinity ($\scri^+$) as $\R_u \times \{ R=0 \} \times S^2_\omega$. The Levi-Civita symbols must be rescaled accordingly~:
\begin{equation} \label{RescEpsilon}
\hat{\varepsilon}_{AB} = R \, \varepsilon_{AB} \, .
\end{equation}
We make on the exterior of the black hole the following choice of unitary (for $\hat{g}$) Newman-Penrose tetrad~:
\begin{equation}
\hat{l}^a \partial_a = -\sqrt\frac{F}{2}\, \partial_R \, ,~ \hat{n}^a \partial_a = \sqrt\frac{2}{F} \left( \partial_u + \frac{R^2 F}{2} \partial_R \right) \, , ~ \hat{m}^a \partial_a = \frac{1}{\sqrt{2}} \left( \partial_\theta + \frac{i}{\sin \theta} \partial_\varphi \right) \, ,  \label{NPTetrad}
\end{equation}
with corresponding dual tetrad
\begin{equation}
\hat{l}_a \d x^a = \sqrt\frac{F}{2}\, \d u \, ,~ \hat{n}_a \d x^a = \sqrt\frac{2}{F} \left( \frac{R^2 F}{2} \d u - \d R \right) \, , ~
\hat{m}_a \d x^a = -\frac{1}{\sqrt{2}} \left( \d \theta + i\sin \theta \d \varphi \right) \, .  \label{NPDualTetrad}
\end{equation}
This is in fact a simple rescaling of a classic Newman-Penrose tetrad for the metric $g$~:
\begin{equation} \label{OriginalTetrad}
l^a \partial_a = \frac{1}{\sqrt{2F}} \left( \partial_t + \partial_{r_*} \right) \, ,~ n^a \partial_a = \frac{1}{\sqrt{2F}} \left( \partial_t - \partial_{r_*} \right) \, ,~ m^a \partial_a = \frac{1}{r\sqrt{2}} \left( \partial_\theta + \frac{i}{\sin \theta} \partial_\varphi \right) \, ,
\end{equation}
since we have
\[ \hat{l}^a = r^2 l^a \, ,~ \hat{n}^a = n^a \, ,~ \hat{m}^a = r m^a \, .\]
The indices for the rescaled frame vectors are lowered with the rescaled metric, so we have the following link with the unrescaled frame co-vectors~:
\[ \hat{l}_a = l_ a \, ,~ \hat{n}_a = R^2 n_a \, ,~ \hat{m}_a = R m_a \, .\]
In terms of associated spin-frames, this corresponds to the rescaling
\begin{equation} \label{RescDyad}
\hat{o}^A = r o^A \, ,~ \hat\iota^A = \iota^A \, ,~ \hat{o}_A = o_A \, ,~ \hat{\iota}_A = R \iota_A \,  .
\end{equation}
We shall use the standard Newman-Penrose notations $D$, $D'$, $\delta$ and $\delta'$ for the directional derivatives $l^a\nabla_a$, $n^a \nabla_a$, $m^a\nabla_a$ and $\bar{m}^a \nabla_a$~; similarly, we denote by $\hat{D}$, $\hat{D}'$, $\hat{\delta}$ and $\hat{\delta}'$ the directional derivatives along $\hat{l}$, $\hat{n}$, $\hat{m}$ and $\bar{\hat{m}}$.

The $4$-volume measure associated with the metric $\hat{g}$ is given by
\begin{equation} \label{4Vol}
\mathrm{dVol}^4 = i \hat{l} \wedge \hat{n} \wedge \hat{m} \wedge \bar{\hat{m}} = - \d u \wedge \d R \wedge \d^2 \omega \, .
\end{equation}
where $\d^2 \omega = i \hat{m}\wedge \bar{\hat{m}}$ is the euclidian measure on the $2$-sphere.

{\bf Note.}  We have denoted by $\hat{l}$, $\hat{n}$, $\hat{m}$ and $\bar{\hat{m}}$ the $1$-forms $\hat{l}_a \d x^a$, $\hat{n}_a \d x^a$, $\hat{m}_a \d x^a$ and $\bar{\hat{m}}_a \d x^a$. We shall use this convention again.

\subsection{Rescaling of spin coefficients}

The rescaling of spin-coefficients under a general conformal rescaling
\[ \hat{g} = \Omega^2 g \, ,~ \hat{o}^A = \Omega^{-1} o^A \, ,~ \hat{\iota}^A = \iota^A \, ,~ \hat{o}_A = o_A \, ,~ \hat{\iota}_A = \Omega \iota_A \, ,\]
is described in \cite{PeRi84} vol. 1 p. 359. The rescaled coefficients are obtained from the original ones by multiplication by a power of $\Omega$ with, for some coefficients, some additional terms that involve the derivatives of $\omega = \log \Omega$ along the original frame vectors. In the special case of the conformal rescaling (\ref{RescMet}), (\ref{RescDyad}), $\Omega =R$ and these terms take the form
\begin{eqnarray*}
D \omega &=& l^a \nabla_a \omega = \frac{1}{\sqrt{2F}} \left( \frac{\partial}{\partial t} + F \frac{\partial}{\partial r} \right) \left( - \log r \right) = -\sqrt{\frac{F}{2}} R \, , \\
\delta ' \omega &=& \bar{m}^a \nabla_a \omega = \frac{1}{\sqrt{2r}} \left( \frac{\partial}{\partial \theta} - \frac{i}{\sin \theta} \frac{\partial}{\partial \varphi} \right) \left( - \log r \right) = 0 \, , \\
\delta  \omega &=& m^a \nabla_a \omega = \frac{1}{\sqrt{2r}} \left( \frac{\partial}{\partial \theta} + \frac{i}{\sin \theta} \frac{\partial}{\partial \varphi} \right) \left( - \log r \right) = 0 \, , \\
D' \omega &=& n^a \nabla_a \omega = \frac{1}{\sqrt{2F}} \left( \frac{\partial}{\partial t} - F \frac{\partial}{\partial r} \right) \left( - \log r \right) = \sqrt{\frac{F}{2}} R \, ,
\end{eqnarray*}
and we have the following relations between the original and rescaled spin-coefficients~:
\[ \begin{array}{|c|c|c|} \hline {\hat{\kappa} = r^3 \kappa} & {\hat{\varepsilon} = r^2 \varepsilon} & {\hat{\pi} = r \pi} \\ \hline {\hat{\rho} = r^2 \rho + \sqrt{\frac{F}{2}} r} & {\hat{\alpha} = r \alpha} & {\hat{\lambda} = \lambda} \\ \hline {\hat{\sigma} = r^2 \sigma} & {\hat{\beta} = r \beta} & {\hat{\mu} = \mu + \sqrt{\frac{F}{2}} R} \\ \hline {\hat{\tau} = r \tau} & {\hat{\gamma} = \gamma - \sqrt{\frac{F}{2}} R} & {\hat{\nu} = R \nu } \\ \hline \end{array} \]
The spin coefficients in the original tetrad (\ref{OriginalTetrad}) have been calculated in \cite{Ni1997}. Using the array above, we obtain~:
\begin{gather}
\hat{\kappa} = \hat{\sigma} = \hat{\lambda} = \hat{\tau} = \hat{\nu} = \hat{\pi} = \hat{\rho} = \hat{\mu} = 0 \, , \nonumber \\
\hat{\varepsilon} = \frac{m}{2 \sqrt{2F}} \, ,~ \hat{\gamma} = \frac{5m R^2 -2R}{2\sqrt{2F}} \, ,~ \hat{\beta} = -\hat{\alpha} = \frac{\cot \theta}
{2 \sqrt{2}} \, . \label{RescSpinCoeff}
\end{gather}
Note that the coefficients $\rho$ and $\mu$ were not zero for the original tetrad\footnote{The property that $\hat\rho =0$ was clear without calculation since $\hat\rho$ represents the geodesic expansion along the flow of $\hat{l}$ (which is a geodesic flow). This is clearly zero since for $\hat{g}$ the surface of the $2$-spheres orthogonal to $\hat{l}$ and $\hat{n}$ is constant. As for $\hat\mu$, it is equal to $-\hat\rho'$, i.e. corresponds to the geodesic contraction along the flow of $\hat{n}$. It is therefore also obviously zero for similar reasons.}.

\subsection{Neighbourhood of spacelike infinity}

We work in the following domain for a given $u_0 << -1$
\[ \Omega_{u_0}^+ := \left\{ (u,R,\omega) \, ;~ u \leq u_0 \, ,~ 0\leq t \leq +\infty \, , ~ \omega \in S^2 \right\} \, .\]
We foliate this neighbourhood of $i^0$ by the hypersurfaces (which are spacelike except for ${\cal H}_0$ which is null)
\[ {\cal H}_{s} = \{ u = -s r_* \, ;~ u \leq u_0 \} \, ,~0 \leq s \leq 1 \, . \]
For $s=1$, the hypersurface ${\cal H}_{1}$ is the part of the $\{ t=0 \}$ surface inside $\Omega_{u_0}^+$ and for $s=0$, ${\cal H}_{0}$ also denoted $\scri^+_{u_0}$ is the part of $\scri^+$ inside $\Omega_{u_0}^+$. The level hypersurfaces of $u$ within $\Omega^+_{u_0}$ will be denoted by ${\cal S}_u$, they are null. Given $0\leq s_1 < s_2 \leq 1$, we will denote by ${\cal S}_u^{s_1,s_2}$ the portion of ${\cal S}_u$ between ${\cal H}_{s_1}$ and ${\cal H}_{s_2}$.

We need an identifying vector field between the hypersurfaces ${\cal H}_{s}$ when decomposing $4$-volume integrals over $\Omega_{u_0}^+$ using the foliation. We use
\begin{equation} \label{VectVHs}
\nu=r_*^2R^2(1-2mR)|u|^{-1}\p_R \, .
\end{equation}
It is tangent to the $u =$ constant surfaces and is naturally associated to the parameter $s$ in that $\nu (s) = 1$. The splitting of the $4$-volume measure $\mathrm{dVol}^4$ corresponding to the foliation $\left\{ { \cal H}_{s} \right\}_{0\leq s \leq 1}$ with identifying vector field $\nu$ is the product of $\d s$ (being the measure along the integral lines of $\nu^a$) and $\nu \hook \mathrm{dVol}^4 |_{{\cal H}_s}= r_*^2 R^2 (1-2mR) |u|^{-1} \d u \d^2\omega |_{{\cal H}_s}$ (which is the resulting $3$-volume measure on each ${\cal H}_s$).

We recall from \cite{MaNi2009} the controls we can infer on $u$, $R$ and $r_*$ in the domain $ \Omega_{u_0}^+$ for $|u_0|$ large enough.
\begin{lemma} \label{ApproxCloseI0}
Let $\varepsilon >0$, then for $u_0<0$, $|u_0|$ large enough, in the domain $\Omega_{u_0}^+$, we have
\[ r< r_* <r(1+\varepsilon ) \, ,~ 1 < Rr_* < 1+\varepsilon  \, ,~ 0 < R|u| < 1+\varepsilon  \, ,~ 1-\varepsilon < 1-2mR <1 \, ,\]
and of course
\[ 0\leq s=\frac{|u|}{r_*} \leq 1 \, . \]
The factor $r_*^2R^2(1-2mR)|u|^{-1}$ appearing in the expression of the vector field $\nu$ satisfies
\[ \frac{1-\varepsilon}{|u|} < r_*^2R^2(1-2mR)|u|^{-1} < \frac{(1+\varepsilon )^2}{|u|} \, .\]
\end{lemma}

\subsection{The Morawetz vector field}

The name ``Morawetz vector field'' is slightly inadequate. It refers to a vector field that is timelike in the neighbourhood of spacelike infinity and is transverse to $\scri^+$. It is constructed from the actual Morawetz vector field in flat spacetime (see \cite{Mo1962}) by expressing it in a coordinate system resembling our $u,R,\omega$ coordinates and brutally keeping the expression on the rescaled Schwarzschild spacetime. More precisely, the Morawetz vector field on Minkowski spacetime is defined by
\[ K = (r^2+t^2) \partial_t + 2tr\partial_r \]
and finds its simplest expression in the coordinates $u=t-r$, $v=t+r$~:
\[ K = u^2 \partial_u + v^2 \partial_v \, .\]
This is a conformal Killing vector field of Minkowski spacetime and is precisely Killing for the Minkowski metric $\eta$ rescaled using the conformal factor $\Omega = 1/r$ (i.e. $\hat\eta =(1/r^2) \eta$). If we use the coordinates $u=t-r$ and $R=1/r$, the vector $K$ takes the form
\[ K = u^2 \partial_u -2 (1+uR) \partial_R \, .\]
We define the ``Morawetz vector field'' on the rescaled Schwarzschild spacetime in the coordinates $u=t-r_*$, $R=1/r$, as
\begin{equation} \label{Morawetz}
T^a \partial_a := u^2 \partial_u -2 (1+uR) \partial_R \, .
\end{equation}
It has the following decomposition on the tetrad $\hat{l}$, $\hat{n}$, $\hat{m}$, $\bar{\hat{m}}$~:
\begin{equation} \label{MorawetzTetrad}
T^a = \sqrt{\frac{2}{F}} \left( 2(1+uR) + \frac{1}{2} (uR)^2 F \right) \hat{l}^a + u^2\sqrt{\frac{F}{2}} \, \hat{n}^a \, .
\end{equation}
Since $K$ is Killing for the rescaled Minkowski metric $R^2 \eta$ and since the Schwarzchild metric is asymptotically flat, the vector field $T$ should provide an approximate Killing vector field, near $i^0$ and $\scri$, for the rescaled metric $\hat{g}$, which is precisely the reason why it was introduced in \cite{MaNi2009} to study the peeling of scalar fields on the Schwarzschild metric. A calculation of its Killing form shows that it is indeed a good approximation of a Killing vector in the vicinity of $i^0$ and $\scri$~:
\[ \nabla_{(a} T_{b)} \d x^a \d x^b = 4mR^2(3+uR)\d u^2 = 8mR^2F^{-1} (3+uR ) \hat{l}_a \hat{l}_b \d x^a \d x^b  \, .\]
Note that a similar construction, based on the null coordinates $(u =t-r_* \, , ~v=t+r_*)$ instead of $(u, R)$, and also referred to as the Morawetz vector field,  was used by Dafermos and Rodnianski in \cite{DaRo}.

\section{Dirac and Maxwell fields} \label{DirMaxFields}

A Dirac spinor field is the direct sum of a neutrino part $\chi^{A'}$ and an anti-neutrino part $\psi_A$~; in the massless case, the two parts decouple and Dirac's equation reduces to the Weyl anti-neutrino equation
\begin{equation} \label{WeylEq}
\nabla^{AA'} \psi_A = 0 \, .
\end{equation}
Similarly, in the source-free case, the anti-self-dual part $\phi_{AB} = \phi_{(AB)}$ and the self-dual part $\bar{\phi}_{A'B'}$ of the electromagnetic field decouple and Maxwell's equations are equivalent to the equations for $\phi_{AB}$~:
\begin{equation} \label{MaxwellEq}
\nabla^{AA'} \phi_{AB} = 0  \, .
\end{equation}
Both equations are conformally invariant~: spinor-valued distributions $\psi_A$ and $\phi_{AB} = \phi_{(AB)}$ satisfy respectively equations \eqref{WeylEq} and \eqref{MaxwellEq} on the exterior of the black hole if and only if the rescaled quantities $\hat{\psi}_A= \Omega^{-1} \psi_A = r \psi_A$ and  $\hat{\phi}_{AB} = \Omega^{-1} \phi_{AB} = r\phi_{AB}$ satisfy on the same domain the rescaled equations
\begin{eqnarray} \label{RescWeylEq}
\hat{\nabla}^{AA'} \hat\psi_A = 0 \, , \\
 \label{RescMaxwellEq}
\hat{\nabla}^{AA'} \hat\phi_{AB} = 0  \, ,
\end{eqnarray}
where $\hat{\nabla}$ is the Levi-Civita connection for the rescaled metric $\hat{g}$.

\subsection{Rescaling of the field components}

We decompose the physical Dirac field $\psi_A$ and Maxwell field $\phi_{AB}$ onto the spin-frame $\{ o^A , \iota^A \}$ and the rescaled fields $\hat{\psi}_A = r\psi_A$ and $\hat{\phi}_{AB}= r\phi_{AB}$ onto the rescaled spin-frame $\{ \hat{o}^A , \hat{\iota}^A \}$. The decomposition is as follows~:
\begin{eqnarray*}
\psi_A &=& \psi_1 o_A -\psi_0 \iota_A \, , \\
\hat{\psi}_A &=& r \psi_A = r \psi_1 o_A - r \psi_0 \iota_A \\
&=& \hat{\psi}_1 \hat{o}_A -\hat{\psi}_0 \hat{\iota}_A = \hat{\psi}_1 o_A -\hat{\psi}_0 R \iota_A\, ,
\end{eqnarray*}
Hence,
\begin{equation} \label{DiracCompRescaled}
\hat{\psi}_0 = r^2 \psi_0 \, ,~\hat{\psi}_1 = r \psi_1 \, .
\end{equation}
A simpler version is the following~:
\[ \hat{\psi}_0 = \hat{\psi}_A \hat{o}^A = r \psi_A r o^A = r^2 \psi_0 \, ,~ \hat{\psi}_1 = \hat{\psi}_A \hat{\iota}^A = r \psi_A \iota^A = r \psi_1 \, .\]
As for the Maxwell field~:
\begin{equation} \label{MaxwellCompRescaled}
\hat{\phi}_0 = r^3 \phi_0 \, ,~ \hat{\phi}_1 = r^2 \phi_1 \, ,~ \hat{\phi}_2 = r \phi_2 \, .
\end{equation}

\subsection{The rescaled equations}

The rescaled Weyl equation \eqref{RescWeylEq} can be expressed using the tetrad (\ref{NPTetrad}) and the associated spin-coefficients as follows (see S. Chandrasekhar \cite{Cha})~:
\[ \left\{ \begin{array}{l}
{ \hat{D}'  \hat{\psi}_0 - \hat{\delta} \hat{\psi}_1 + (\hat{\mu} - \hat{\gamma} ) \hat{\psi}_0 + (\hat{\tau} - \hat{\beta} )
\hat{\psi}_1 = 0 \, , } \\ \\
{ \hat{D} \hat{\psi}_1 - \hat{\delta}' \hat{\psi}_0 + (\hat{\alpha} - \hat{\pi} )\hat{\psi}_0 + (\hat{\varepsilon} -
\hat{\rho} ) \hat{\psi}_1 = 0 \, ,}  \end{array} \right. \]
the link being
\begin{eqnarray}
0 = \hat{\nabla}^{AA'} \hat{\psi}_A &=& \left( \hat{D}' \hat{\psi}_0 - \hat{\delta} \hat{\psi}_1 + (\hat{\mu} - \hat{\gamma} ) \hat{\psi}_0 + (\hat{\tau} - \hat{\beta} )
\hat{\psi}_1 \right) \bar{\hat{o}}^{A'} \nonumber \\
&& +  \left( \hat{D} \hat{\psi}_1 - \hat{\delta}' \hat{\psi}_0 + (\hat{\alpha} - \hat{\pi} )\hat{\psi}_0 + (\hat{\varepsilon} -
\hat{\rho} ) \hat{\psi}_1 \right) \bar{\hat\iota}^{A'} \, . \label{RescWeylEqComponents}
\end{eqnarray}
This gives us the system (~with $F = 1-2mR)$:
\[ \left\{ \begin{array}{l}
{ \sqrt{\frac{2}{F}} \left( \partial_u + \frac{1}{2} R^2 F \partial_R \right) \, \hat{\psi}_0 - \frac{1}{\sqrt{2}} \left( \partial_\theta + \frac{1}{2} \cot \theta + \frac{i}{\sin \theta} \partial_\varphi \right) \, \hat{\psi}_1 - \frac{5m R^2 -2R}{2\sqrt{2F}} \hat{\psi}_0  = 0 \, , } \\ \\
{ -\sqrt{\frac{F}{2}} \partial_R \hat{\psi}_1 -  \frac{1}{\sqrt{2}} \left( \partial_\theta + \frac{1}{2} \cot \theta - \frac{i}{\sin \theta} \partial_\varphi \right)  \, \hat{\psi}_0 + \frac{m}{2 \sqrt{2F}} \hat{\psi}_1 = 0 \, .}  \end{array} \right. \]
This can be simplified as follows~:
\begin{equation} \label{WeylEqGHP}
\left\{ \begin{array}{l}
{ \hat\thorn' \hat{\psi}_0 - \hat{\eth}  \hat{\psi}_1 = 0 \, , } \\ \\
{ \hat\thorn \hat{\psi}_1 - \hat{\eth}' \hat{\psi}_0 = 0 \, ,}  \end{array} \right.
\end{equation}
where $\hat\thorn$, $\hat\thorn'$, $\hat{\eth}$ and $\hat{\eth}'$ are the weighted differential operators of the GHP formalism (Geroch-Held-Penrose \cite{GHP}, also referred to as compacted spin-coefficient formalism in Penrose and Rindler \cite{PeRi84} Vol.1 section 4.12), which, applied to $\hat{\psi}_0$ and $\hat{\psi}_1$ take the form
\begin{eqnarray*}
\hat\thorn' \hat{\psi}_0 = \sqrt{\frac{2}{F}} \left( \partial_u + \frac{R^2 F}{2} \partial_R + \frac{2R-5m R^2}{4} \right) \, \hat{\psi}_0 &,& \hat\thorn \hat{\psi}_1 = -\sqrt{\frac{F}{2}} \left( \partial_R - \frac{m}{2F} \right) \hat{\psi}_1 \, ,\\
\hat{\eth} \hat{\psi}_1 = \frac{1}{\sqrt{2}} \left( \partial_\theta + \frac{1}{2} \cot \theta + \frac{i}{\sin \theta} \partial_\varphi \right) \, \hat{\psi}_1 &,& \hat{\eth}' \hat{\psi}_0 = \frac{1}{\sqrt{2}} \left( \partial_\theta + \frac{1}{2} \cot \theta - \frac{i}{\sin \theta} \partial_\varphi \right)  \, \hat{\psi}_0 \, .
\end{eqnarray*}
For the rescaled anti-self-dual Maxwell system, we have
\begin{eqnarray*}
0=\nabla^{AA'} \hat\phi_{AB} &=& \left(  \hat{D}' \hat{\phi}_0 - \hat{\delta} \hat{\phi}_1 + (\hat{\mu} - 2\hat{\gamma} ) \hat{\phi}_0 + 2 \hat{\tau} 
\hat{\phi}_1 - \hat\sigma \hat\phi_2 \right) \bar{\hat{o}}^{A'} \hat{o}_B \\
&&- \left( \hat{D} \hat{\phi}_1 -\hat{\delta}' \hat{\phi}_0 + (2 \hat{\alpha} - \hat{\pi} )\hat{\phi}_0 - 2 \hat{\rho} \hat{\phi}_1 + \hat\kappa \hat\phi_2 \right) \bar{\hat{o}}^{A'} \hat{\iota}_B  \\
&& + \left( \hat{D}' \hat{\phi}_1 - \hat{\delta} \hat{\phi}_2 - \hat{\nu} \hat{\phi}_0 + 2\hat{\mu} \hat{\phi}_1 + (\hat\tau - 2\hat\beta ) \hat\phi_2 \right) \bar{\hat{\iota}}^{A'} \hat{o}_B \\
&& - \left( \hat{D} \hat{\phi}_2 - \hat{\delta}' \hat{\phi}_1 + \hat{\lambda} \hat{\phi}_0 - 2 \hat\pi \hat\phi_1 + (2 \hat{\varepsilon} - \hat{\rho} ) \hat{\phi}_2 \right) \bar{\hat{\iota}}^{A'} \hat{\iota}_B \, ,
\end{eqnarray*}
so equation \eqref{RescMaxwellEq} is equivalent to the system
\[ \left\{ \begin{array}{l}
{ \hat{D}' \hat{\phi}_0 - \hat{\delta} \hat{\phi}_1 + (\hat{\mu} - 2\hat{\gamma} ) \hat{\phi}_0 + 2 \hat{\tau} 
\hat{\phi}_1 - \hat\sigma \hat\phi_2 = 0 \, , } \\ \\
{ \hat{D} \hat{\phi}_1 - \hat{\delta}' \hat{\phi}_0 + (2 \hat{\alpha} - \hat{\pi} )\hat{\phi}_0 - 2 \hat{\rho} \hat{\phi}_1 + \hat\kappa \hat\phi_2 = 0 \, ,} \\ \\
{ \hat{D}' \hat{\phi}_1 - \hat{\delta} \hat{\phi}_2 - \hat{\nu} \hat{\phi}_0 + 2\hat{\mu} \hat{\phi}_1 + (\hat\tau - 2\hat\beta ) \hat\phi_2 = 0 \, , } \\ \\
{ \hat{D} \hat{\phi}_2 - \hat{\delta}' \hat{\phi}_1 + \hat{\lambda} \hat{\phi}_0 - 2 \hat\pi \hat\phi_1 + (2 \hat{\varepsilon} - \hat{\rho} ) \hat{\phi}_2 = 0 \, . }   \end{array} \right. \]
In the Geroch-Held-Penrose formalism, this takes on the simpler expression
\begin{equation} \label{GHPMaxwell}
\left\{ \begin{array}{l}
{ \hat\thorn ' \hat{\phi}_0 - \hat{\eth} \hat{\phi}_1 = 0 \, , } \\ \\
{ \hat\thorn \hat{\phi}_1 - \hat{\eth}' \hat{\phi}_0 = 0 \, ,} \\ \\
{ \hat\thorn' \hat{\phi}_1 - \hat{\eth} \hat{\phi}_2 = 0 \, , } \\ \\
{ \hat\thorn \hat{\phi}_2 - \hat{\eth}' \hat{\phi}_1 = 0 \, ,}   \end{array} \right.
\end{equation}
with
\begin{eqnarray*}
\hat\thorn ' \hat\phi_0 = \sqrt{\frac{2}{F}} ( \partial_u + \frac{1}{2} R^2F \partial_R ) \hat\phi_0 -\frac{5mR^2 -2R}{\sqrt{2F}} \hat{\phi}_0  &,& \hat\thorn \hat\phi_1 = -\sqrt{\frac{F}{2}} \partial_R \hat{\phi}_1 \, ,\\
\hat\thorn ' \hat\phi_1 = \sqrt{\frac{2}{F}} ( \partial_u + \frac{1}{2} R^2F \partial_R ) \hat{\phi}_1 &,& \hat\thorn \hat\phi_2 = -\sqrt{\frac{F}{2}} \partial_R \hat{\phi}_2 + \frac{m}{\sqrt{2F}} \hat{\phi}_2 \, , \\
\hat{\eth} \hat\phi_1 = \frac{1}{\sqrt{2}} ( \partial_\theta + \frac{i}{\sin \theta} \partial_\varphi ) \hat{\phi}_1 &,& \hat{\eth}' \hat\phi_0 = \frac{1}{\sqrt{2}} ( \partial_\theta + \cot \theta - \frac{i}{\sin \theta} \partial_\varphi ) \hat{\phi}_0 \, , \\
\hat{\eth} \hat\phi_2 = \frac{1}{\sqrt{2}} ( \partial_\theta + \cot \theta + \frac{i}{\sin \theta} \partial_\varphi ) \hat{\phi}_2 &,& \hat{\eth}' \hat\phi_1 = \frac{1}{\sqrt{2}} ( \partial_\theta - \frac{i}{\sin \theta} \partial_\varphi ) \hat{\phi}_1 \, .
\end{eqnarray*}

\subsection{Conserved quantity and estimates for perturbed equations}

\subsubsection{The Weyl equation}

The conserved current for the Weyl equation is $J^a = \hat{\psi}^A \bar{\hat{\psi}}^{A'}$, which gives the following closed $3$-form by contraction with the $4$-volume measure \eqref{4Vol}~:
\begin{eqnarray}
\omega &:=& * J_a \d x^a = J \hook \dvol^4\nonumber \\
& = & \left( \left| \hat{\psi}_1 \right|^2 \hat{l}^a \partial_a  + \left| \hat{\psi}_0 \right|^2 \hat{n}^a \partial_a - \hat{\psi}_1 \overline{\hat{\psi}_0} \hat{m}^a \partial_a - \overline{\hat{\psi}_1} \hat{\psi}_0 \bar{\hat{m}}^a \partial_a \right) \hook \dvol^4 \nonumber \\
&=& -\vert \hat\psi_1 \vert^2 \hat{l} \wedge \d^2 \omega + \vert \hat\psi_0 \vert^2 \hat{n} \wedge \d^2 \omega - i \hat{\psi}_1 \overline{\hat{\psi}_0} \, \hat{l} \wedge \hat{n} \wedge \hat{m} + i \hat{\psi}_0 \overline{\hat{\psi}_1} \, \hat{l} \wedge \hat{n} \wedge \bar{\hat{m}}  \nonumber \\
&=& -\sqrt{\frac{2}{F}} \left| \hat{\psi}_0 \right|^2 \d R \wedge \d^2 \omega - \sqrt{\frac{F}{2}} \left( \left| \hat{\psi}_1 \right|^2 - R^2 \left| \hat{\psi}_0 \right|^2 \right) \d u \wedge \d^2 \omega \nonumber \\
&& + \sqrt{2} \Re \left( \hat{\psi}_0 \overline{\hat{\psi}_1}\right) \d u \wedge \d R \wedge \sin \theta \d \varphi + \sqrt{2} \Re \left( i \hat{\psi}_0 \overline{\hat{\psi}_1}\right) \d u \wedge \d R \wedge \d \theta \, . \label{DiracClosed3Form}
\end{eqnarray}
On a given hypersurface ${\cal H}_{s}$, we have
\[ \d R = \frac{FR^2}{s} \d u \]
and the conserved quantity takes the form
\begin{eqnarray}
{\cal E}_{{\cal H}_s} (\hat{\psi} ) := \int_{{\cal H}_s} \omega &=& \int_{{\cal H}_s} \left( \left( \frac{2}{s} - 1 \right) R^2 \left| \hat{\psi}_0 \right|^2 + \left| \hat{\psi}_1 \right|^2 \right)  \sqrt{\frac{F}{2}} \d u \, \d^2 \omega \nonumber \\
&=& \int_{{\cal H}_s} \left( \left( \frac{2r_*}{|u|} - 1 \right) R^2 \left| \hat{\psi}_0 \right|^2 + \left| \hat{\psi}_1 \right|^2 \right)  \sqrt{\frac{F}{2}} \d u \, \d^2 \omega \, .
\end{eqnarray}
On ${\cal S}_u$,
\begin{eqnarray}
{\cal E}_{{\cal S}_{u}} (\hat{\psi} ) := \int_{{\cal S}_{u}} \omega &=& \int_{{\cal S}_{u}} \sqrt{\frac{2}{F}} \left| \hat{\psi}_0 \right|^2 \d R \, \d^2 \omega \, .
\end{eqnarray}
\begin{lemma}
The energies on ${\cal S}_u$, $u\leq u_0$ and ${\cal H}_s$, $0 \leq s \leq 1$ have the following simpler equivalents (meaning that there are constants independent of $u\leq u_0$, $0\leq s \leq 1$ and the smooth spinor field $\hat{\psi}_A$ such that the energies on ${\cal S}_u$ and ${\cal H}_s$ are controlled above and below by these constants times the simpler expressions)~:
\begin{eqnarray}
{\cal E}_{{\cal S}_{u}} (\hat{\psi} ) &\simeq & \int_{{\cal S}_{u}} \left| \hat{\psi}_0 \right|^2 \d R \, \d^2 \omega \, , \\
{\cal E}_{{\cal H}_s} (\hat{\psi} ) &\simeq & \int_{]-\infty , u_0 [_u \times S^2_\omega} \left( \frac{R}{\vert u \vert} \left| \hat{\psi}_0 \right|^2 + \left| \hat{\psi}_1 \right|^2 \right) \d u \, \d^2 \omega \, .
\end{eqnarray}
\end{lemma}
{\bf Proof.} This is a direct consequence of lemma \ref{ApproxCloseI0} and of the fact that
\[ \frac{1}{s} \leq \frac{2}{s} -1 \leq \frac{2}{s} \, . \qed \]

The closedness of the $3$-form $\omega$ gives for any smooth solution $\hat{\psi}_A$ of \eqref{RescWeylEq} with compactly supported initial data~:
\begin{equation} \label{L2EnEq}
{\cal E}_{{\cal H}_{s_1}} (\hat{\psi} ) + {\cal E}_{{\cal S}_{u_0}^{s_1,s_2}} (\hat{\psi} ) = {\cal E}_{{\cal H}_{s_2}} (\hat{\psi} ) \mbox{ for any } 0\leq s_1<s_2\leq 1\, .
\end{equation}
Now, consider a Dirac equation with error terms of two types, a potential $P$ and a source $Q$~:
\begin{equation} \label{DiracErrorTerm}
\nabla^{AA'} \hat{\psi}_A = P^{AA'} \hat{\psi}_A + Q^{A'} \, .
\end{equation}
Then, differentiating the current $1$-form $J_a$, we get
\[ \nabla^{AA'} \left( \hat{\psi}_A \bar{\hat{\psi}}_{A'} \right) = \left( P^{AA'} + \bar{P}^{AA'} \right) \hat{\psi}_A \bar{\hat{\psi}}_{A'} + 2 \Re \left( Q^{A'} \bar{\hat{\psi}}_{A'} \right)  \]
which is an approximate conservation law. When integrating this over the $4$-volume $\Omega_{u_0}^{s_1 ,s_2}$ bounded by hypersurfaces  ${\cal H}_{s_1}$, ${\cal H}_{s_2}$, for $0\leq s_1<s_2\leq 1$, and the part ${\cal S}^{s_1,s_2}_{u_0}$ of ${\cal S}_{u_0}$ between ${\cal H}_{s_1}$ and ${\cal H}_{s_2}$, we obtain for any smooth solution of \eqref{DiracErrorTerm} with compactly supported initial data~:
\begin{gather*}
\left\vert {\cal E}_{{\cal H}_{s_1}} (\hat{\psi} ) - {\cal E}_{{\cal H}_{s_2}} (\hat{\psi} ) + {\cal E}_{{\cal S}_{u_0}^{s_1,s_2}} (\hat{\psi} ) \right\vert = \left\vert \int_{\Omega_{u_0}^{s_1 ,s_2}} \left( \left( P^{AA'} + \bar{P}^{AA'} \right) \hat{\psi}_A \bar{\hat{\psi}}_{A'} + 2 \Re \left( Q^{A'} \bar{\hat{\psi}}_{A'} \right) \right) \dvol^4 \right\vert \\
\leq (1+\varepsilon )^2 \int_{s_1}^{s_2} \int_{{\cal H}_s} \left\vert \left( P^{AA'} + \bar{P}^{AA'} \right) \hat{\psi}_A \bar{\hat{\psi}}_{A'} + 2 \Re \left( Q^{A'} \bar{\hat{\psi}}_{A'} \right) \right\vert \frac{1}{\vert u \vert} \d u \d^2 \omega \d s \, .
\end{gather*}
Energy estimates will be established using the Gronwall inequality provided the integrand on the right-hand side can be estimated by the energy density on ${\cal H}_s$, in a sufficiently uniform way so as not to prevent integrability in $s$.

\subsubsection{The Maxwell system}

An anti-self-dual Maxwell field $\hat{\phi}_{AB}$ has a stress-energy tensor given by the following expression
\[ T_{ab} = \hat{\phi}_{AB} \bar{\hat{\phi}}_{A'B'} \, .\]
In order to define an energy current, we need to choose a timelike vector field to contract the stress-energy tensor with. A natural timelike vector field would be the Killing vector $\partial_u$ which is equal to $\partial_t$ in the Schwarzschild coordinate system, but $\partial_u$ becomes null on $\scri^+$ and we require more control there in order to establish peeling results. So we use the Morawetz vector field \eqref{Morawetz}
\[ T^a \partial_a = u^2 \partial_u - 2(1+uR ) \partial_R \, .\]
The associated energy current is the vector field $V^a = T^{ab} T_b$ whose decomposition on the Newman-Penrose tetrad $\hat{l}, \hat{n}, \hat{m}, \bar{\hat{m}}$ is given by
\begin{eqnarray*}
V^a &=& \left( \sqrt{\frac{2}{F}} (2 + 2uR + \frac{(uR)^2}{2} F ) \vert \hat\phi_1 \vert^2 + \sqrt{\frac{F}{2}} u^2 \vert \hat\phi_2 \vert^2 \right) \hat{l}^a \\
&& + \left( \sqrt{\frac{2}{F}} (2 + 2uR + \frac{(uR)^2}{2} F ) \vert \hat\phi_0 \vert^2 + \sqrt{\frac{F}{2}} u^2 \vert \hat\phi_1 \vert^2 \right) \hat{n}^a \\
&& - \left( \sqrt{\frac{2}{F}} (2 + 2uR + \frac{(uR)^2}{2} F ) \hat\phi_1 \overline{\hat{\phi}_0} + \sqrt{\frac{F}{2}} u^2 \hat\phi_2 \overline{\hat{\phi}_1} \right) \hat{m}^a \\
&& - \left( \sqrt{\frac{2}{F}} (2 + 2uR + \frac{(uR)^2}{2} F ) \hat\phi_0 \overline{\hat{\phi}_1} + \sqrt{\frac{F}{2}} u^2 \hat\phi_1 \overline{\hat{\phi}_2} \right) \hat{\bar{m}}^a \, .
\end{eqnarray*}
Since $T^a$ is not an exact Killing vector, $V^a$ is not divergence free and it satisfies merely an approximate conservation law
\begin{equation} \label{MaxwellErrorTerm}
\nabla_a V^a = \nabla_{(a} T_{b)} T^{ab} = 8mR^2 F^{-1} (3+uR) T_{ab} l^a l^b = 8mR^2 F^{-1} (3+uR) \vert \hat{\phi}_0 \vert^2 \, .
\end{equation}
The energy $3$-form is the Hodge dual of the energy current
\begin{eqnarray*}
\omega &:=& * (V_a \d x^a ) = V \lrcorner \dvol^4 \\
&= & \left( \sqrt{\frac{2}{F}} (2 + 2uR + \frac{(uR)^2}{2} F ) \vert \hat\phi_1 \vert^2 + \sqrt{\frac{F}{2}} u^2 \vert \hat\phi_2 \vert^2 \right)  (-\hat{l} \wedge \d^2 \omega ) \\
&& + \left( \sqrt{\frac{2}{F}} (2 + 2uR + \frac{(uR)^2}{2} F ) \vert \hat\phi_0 \vert^2 + \sqrt{\frac{F}{2}} u^2 \vert \hat\phi_1 \vert^2 \right) (\hat{n} \wedge \d^2 \omega ) \\
&& + \left( \sqrt{\frac{2}{F}} (2 + 2uR + \frac{(uR)^2}{2} F ) \hat\phi_1 \overline{\hat{\phi}_0} + \sqrt{\frac{F}{2}} u^2 \hat\phi_2 \overline{\hat{\phi}_1} \right) (-i\hat{l} \wedge \hat{n} \wedge \hat{m}) \\
&& + \left( \sqrt{\frac{2}{F}} (2 + 2uR + \frac{(uR)^2}{2} F ) \hat\phi_0 \overline{\hat{\phi}_1} + \sqrt{\frac{F}{2}} u^2 \hat\phi_1 \overline{\hat{\phi}_2} \right) (i\hat{l} \wedge \hat{n} \wedge \bar{\hat{m}}) \, ,
\end{eqnarray*}
(recall that $\d^2 \omega = i \hat{m} \wedge \bar{\hat{m}}$ is the euclidian measure on $S^2$).

On a $u=$constant hypersurface ${\cal S}_u$, the energy is given by
\begin{equation} \label{MEnergySu}
{\cal E}_{{\cal S}_u} (\hat{\phi} ) := \int_{{\cal S}_u} \omega = \int_{{\cal S}_u} \left( \sqrt{\frac{2}{F}} (2 + 2uR + \frac{(uR)^2}{2} F ) \vert \hat\phi_0 \vert^2 + \sqrt{\frac{F}{2}} u^2 \vert \hat\phi_1 \vert^2 \right) \d R \, \d^2 \omega \, .
\end{equation}
On a $u=-sr_*$ hypersurface ${\cal H}_s$, recall that
\[ \d R = \frac{FR^2}{s}\d u \]
and therefore (for $0<s\leq 1$)
\begin{eqnarray}
{\cal E}_{{\cal H}_s} (\hat{\phi} ) := \int_{{\cal H}_s} \omega &=& \int_{{\cal H}_s} \left( \left( \frac{2}{s}-1 \right) R^2 \left( 2+2uR + \frac{(uR)^2}{2} F \right) \vert \hat{\phi}_0 \vert^2 \right. \nonumber \\
&& \hspace{0.5in} \left. + \left( 2 + 2uR + \frac{(uR)^2F}{s} \right) \vert \hat{\phi}_1 \vert^2 + \frac{F}{2} u^2 \vert \hat{\phi}_2 \vert^2 \right) \d u \, \d^2 \omega \, .
\label{MEnergyHs}
\end{eqnarray}
As $s\rightarrow 0$, this expression simplifies to give the energy on $\scri^+$
\begin{equation} \label{MEnergyScriPlus}
{\cal E}_{\scri^+_{u_0}} (\hat{\phi} ) := \int_{\scri^+_{u_0}} \omega = \int_{\scri^+_{u_0}} \left( 2 \vert \hat{\phi}_1 \vert^2 + \frac{u^2}{2} \vert \hat{\phi}_2 \vert^2 \right) \d u \, \d^2 \omega \, ,
\end{equation}
using the fact that
\[ \frac{(uR)^2}{s} = (-uR) Rr_* \rightarrow 0 \mbox{ as } r \rightarrow +\infty \mbox{ with } u \mbox{ bounded } \]
and similarly $R^2 /s \rightarrow 0$ as $r\rightarrow +\infty$.
\begin{lemma}
We have the following equivalent simpler expressions fo the energies on ${\cal S}_u$, $u\leq u_0$ and ${\cal H}_s$, $0\leq s \leq 1$
\begin{eqnarray}
{\cal E}_{{\cal S}_u} (\hat{\phi} ) & \simeq & \int_{{\cal S}_u} \left(  \vert \hat\phi_0 \vert^2 + u^2 \vert \hat\phi_1 \vert^2 \right) \d R \, \d^2 \omega \, , \\
{\cal E}_{{\cal H}_s} (\hat{\phi} )  &\simeq & \int_{{\cal H}_s} \left(  \frac{R}{\vert u \vert} \vert \hat{\phi}_0 \vert^2 +  \vert \hat{\phi}_1 \vert^2 + u^2 \vert \hat{\phi}_2 \vert^2 \right) \d u \, \d^2 \omega \, .
\end{eqnarray}
\end{lemma}
{\bf Proof.} We notice that $2+2uR + \frac{(uR)^2}{2} F $ vanishes for
\[ uR = -\frac{2}{F} ( 1 \pm \sqrt{2mR} ) \, .\]
We know from lemma \ref{ApproxCloseI0} that in $\Omega^+_{u_0}$, $-1-\varepsilon < uR \leq 0$ with $0\leq \varepsilon <<1$ and also
\[2(1-\varepsilon ) < \frac{2}{F} ( 1 \pm \sqrt{2mR} ) < 2 (1+\varepsilon ) \, .\]
It follows that $2+2uR + \frac{(uR)^2}{2} F$ vanishes nowhere for $0< s \leq 1$ and tends to $2$ as $uR \rightarrow 0$, so this quantity is also bounded below away from zero and above, uniformly on $\Omega^+_{u_0}$. The lemma then follows from lemma \ref{ApproxCloseI0}. \qed

\section{Peeling} \label{Peeling}

\subsection{Peeling for Dirac}

We have already established energy estimates for $\hat\psi_A$ between $\scri^+_{u_0}$, ${\cal S}_{u_0}$ and ${\cal H}_1$. Now we establish estimates for successive derivatives of $\hat\psi_A$. We do not need to commute all directional derivatives into the equation~; as our goal is to control transverse regularity on $\scri^+$ we focus on derivatives in the direction of $\partial_R$, i.e. of $\hat{l}^a$. We denote by $D_R \hat{\psi}_A$ the spinor
\[ D_R \hat{\psi}_A := \partial_R \hat{\psi}_1 o_A - \partial_R \hat{\psi}_0 \iota_A \]
and we work out the equation satisfied by $D_R \hat{\psi}$. In order to obtain a more useable expression for this equation we multiply the first line of \eqref{WeylEqGHP} by $\sqrt{2F}$, keep the second as it is and commute $\partial_R$ into the resulting system. We get~:
\[ \left\{ \begin{array}{l}
{ \left( 2 \partial_u + R^2 F \partial_R \right) \, \partial_R \hat{\psi}_0 - \sqrt{2F} \, \hat{\eth}  \partial_R \hat{\psi}_1 + R \left( 1- \frac{5mR}{2} \right) \partial_R \hat{\psi}_0  } \\
{ \hspace{1in} = -2R \left( 1-3mR \right) \partial_R \hat{\psi}_0 - m \sqrt{\frac{2}{F}} \hat{\eth} \hat{\psi}_1 - \left( 1- 5mR \right)\hat{\psi}_0 \, ,} \\ \\
{ -\sqrt{\frac{F}{2}} \left( \partial_R - \frac{m}{2F} \right) \partial_R \hat{\psi}_1 - \hat{\eth}' \partial_R \hat{\psi}_0 = -\frac{m}{\sqrt{2F}} \partial_R \hat{\psi}_1 - \frac{m^2}{(2F)^{3/2}} \hat{\psi}_1 \, .}  \end{array} \right. \]
This can be re-written as
\[ \left\{ \begin{array}{l}
{ \sqrt{\frac{2}{F}} \left( \partial_u + \frac{R^2 F}{2} \partial_R + \frac{2R-5mR^2}{4} \right) \, \partial_R \hat{\psi}_0 - \hat{\eth}  \partial_R \hat{\psi}_1 }  \\
{ \hspace{1in} =  -\sqrt{\frac{2}{F}} R (1-3mR) \partial_R \hat{\psi}_0 - \frac{m}{F} \hat{\eth} \hat{\psi}_1 - \frac{1- 5mR}{\sqrt{2F}} \hat{\psi}_0 \, ,} \\ \\
{ -\sqrt{\frac{F}{2}} \left( \partial_R - \frac{m}{2F} \right) \partial_R \hat{\psi}_1 - \hat{\eth}' \partial_R \hat{\psi}_0 = -\frac{m}{\sqrt{2F}} \partial_R \hat{\psi}_1 - \frac{m^2}{(2F)^{3/2}} \hat{\psi}_1}  \end{array} \right. \]
and as a spinorial equation takes the form
\begin{eqnarray}
\hat{\nabla}^{AA'} \left( D_R \hat{\psi}_A \right) &=& - \left( \sqrt{\frac{2}{F}} R (1-3mR) \partial_R \hat{\psi}_0  + \frac{m}{F} \hat{\eth} \hat{\psi}_1 + \frac{1- 5mR}{\sqrt{2F}} \hat{\psi}_0 \right) o^{A'} \nonumber \\
&& - \left( \frac{m}{\sqrt{2F}} \partial_R \hat{\psi}_1 + \frac{m^2}{(2F)^{3/2}} \hat{\psi}_1 \right) \iota^{A'} \, .\label{WeylEqDrPsi}
\end{eqnarray}
The conservation law associated with equation \eqref{WeylEqDrPsi} is the following
\begin{eqnarray*}
\hat{\nabla}^{AA'} \left[ \left( D_R \hat{\psi}_A \right) \left( D_R \bar{\hat{\psi}}_{A'} \right) \right] &=& 2 \Re \left[ \left( \hat{\nabla}^{AA'} D_R \hat{\psi}_A \right) D_R \bar{\hat{\psi}}_{A'} \right] \\
&=& - 2 \Re \left[ \left( \sqrt{\frac{2}{F}} R (1-3mR) \partial_R \hat{\psi}_0 + \frac{m}{F} \hat{\eth} \hat{\psi}_1 + \frac{1- 5mR}{\sqrt{2F}} \hat{\psi}_0 \right) \overline{\partial_R \hat{\psi}_0} \right. \\
&& \left. + \left( \frac{m}{\sqrt{2F}} \partial_R \hat{\psi}_1 + \frac{m^2}{(2F)^{3/2}} \hat{\psi}_1 \right) \overline{\partial_R \hat{\psi}_1} \right]
\end{eqnarray*}
In order to obtain energy estimates for $D_R \hat{\psi}$ using Gronwall's inequality, we need to estimate the right-hand side by the energy densities for $\hat{\psi}$ or $D_R \hat{\psi}$. Two types of terms present a difficulty~: those involving angular derivatives and the term involving $ \hat{\psi}_0 $ and $\partial_R \hat{\psi}_0 $ without a factor of $R$. In order to control the angular terms, we must commute angular derivatives into the equation~; this will turn out to give us the additional control on $\hat{\psi}_0$ that we need.

The commutation of angular derivatives into the Weyl equation is best described, and performed, using the GHP formalism, i.e. we use the form \eqref{WeylEqGHP} of the Weyl equation. In our framework, using the values of the spin coefficients, we have the following identities
\begin{equation} \label{GHPCommutators}
\left[ \hat{\thorn} \, ,~\hat{\eth} \right] \eta =0 \mbox{ for } \eta \mbox{ of weight } \{ -1, 0 \} \mbox{ and } \left[ \hat{\thorn}' \, ,~\hat{\eth}' \right] \eta =0 \mbox{ for } \eta \mbox{ of weight } \{ 1, 0 \} \, .
\end{equation}
So when we apply $\eth'$ to the first equation and $\eth$ to the second, we obtain
\begin{equation} \label{WeylEqEthPsi}
\left\{ \begin{array}{l}
{ \hat{\thorn} ' \hat{\eth}' \hat{\psi}_0 - \hat{\eth}' \hat{\eth} \hat{\psi}_1 =0  } \\
{ \hat{\thorn} \hat{\eth} \hat{\psi}_1- \hat{\eth} \hat{\eth}' \hat{\psi}_0 =0 \, .}  \end{array} \right.
\end{equation}
Note that $\hat{\eth}' \hat{\psi}_0$ has weight $\{ 0,1 \}$ and $\hat{\eth} \hat{\psi}_1$ weight $\{ 0, -1 \}$, so putting
\[ (D_\omega \hat{\psi} )_{A'} = \hat{\eth} \hat{\psi}_1 \hat{o}_{A'} - \hat{\eth}' \hat{\psi}_0 \hat{\iota}_{A'} \, ,\]
equation \eqref{WeylEqEthPsi} can be written as a conjugate Weyl equation
\[ \nabla^{AA'} (D_\omega \hat{\psi} )_{A'} =0 \, , \]
i.e. $D_\omega$ is a symmetry operator for the Dirac equation on the rescaled Schwarzchild metric, sending anti-neutrino fields to neutrino fields.

Now consider some smooth solution $\hat{\psi}_A$ of \eqref{RescWeylEq} with compactly supported initial data. Both $\hat{\psi}$ and $D_\omega \hat{\psi}$ satisfy the energy equality \eqref{L2EnEq}, which entails for any $s \in [0,1]$~:
\begin{eqnarray*}
\int_{{\cal H}_1} \left( \frac{R}{\vert u \vert} \vert \hat{\psi}_0 \vert^2 + \vert \hat{\psi}_1 \vert^2 \right) \d u \d^2 \omega & \simeq & \int_{{\cal H}_s} \left( \frac{R}{\vert u \vert} \vert \hat{\psi}_0 \vert^2 + \vert \hat{\psi}_1 \vert^2 \right) \d u \d^2 \omega \\
&&+ \int_{{\cal S}_{u}^{s,1}} \left| \hat{\psi}_0 \right|^2 \d R \d^2 \omega\, , \\
\int_{{\cal H}_1} \left( \frac{R}{\vert u \vert} \vert \hat{\eth}' \hat{\psi}_{0} \vert^2 + \vert \hat{\eth} \hat{\psi}_{1} \vert^2 \right) \d u \d^2 \omega & \simeq & \int_{{\cal H}_s} \left( \frac{R}{\vert u \vert} \vert \hat{\eth}' \hat{\psi}_{0} \vert^2 + \vert \hat{\eth} \hat{\psi}_{1} \vert^2 \right) \d u \d^2 \omega \\
&&+ \int_{{\cal S}_{u}^{s,1}} \left| \hat{\eth}' \hat{\psi}_{0} \right|^2 \d R \d^2 \omega\, . \\
\end{eqnarray*}
Note that this immediately gives us a control of the $4$-volume $L^2$ norm of $\hat{\psi}_0$ and $\hat{\eth}' \hat{\psi}_0$ in terms of the energy on ${\cal H}_1$ using the foliation by ${\cal S}_u$ with the identifying vector field $\partial_u$. However, this works only once we have obtained the estimates, it does not allow to control a perturbed equation with an error term $( \hat{\psi}_0)^2$ in the conservation law.

Integrating on $\Omega^{s_1,s_2}_{u_0}$ the conservation law associated with equation \eqref{WeylEqDrPsi}, we obtain
\begin{gather*}
\left\vert {\cal E}_{{\cal H}_{s_1}} (D_R \hat{\psi}) + {\cal E}_{{\cal S}^{s_1,s_2}_{u_0}} (D_R \hat{\psi}) - {\cal E}_{{\cal H}_{s_2}} (D_R \hat{\psi}) \right\vert \\
\leq 2\int_{s_1}^{s_2} \int_{{\cal H}_{s}} \left( \left\vert -\sqrt{\frac{2}{F}} R (1-3mR) \partial_R \hat{\psi}_0 - \frac{m}{F} \hat{\eth} \hat{\psi}_1 - \frac{1- 5mR}{\sqrt{2F}} \hat{\psi}_0 \right\vert \vert \partial_R \hat{\psi}_0\vert \right. \\
\left. + \left\vert -\frac{m}{\sqrt{2F}} \partial_R \hat{\psi}_1 - \frac{m^2}{2F\sqrt{2F}} \hat{\psi}_1 \right\vert \vert \partial_R \hat{\psi}_1\vert \right) \frac{1}{\vert u \vert} \d u \d^2 \omega \d s
\end{gather*}
The last two error terms are trivially controlled by the energies of $\hat{\psi}_A$ and $D_R \hat{\psi}_A$~; the first error term, thanks to the $1/\vert u \vert$ coming from the Leray measure, is exactly controlled by the energy of $D_R \hat{\psi}_A$. The difficulties are with the second and third error terms. For the second term, we use the control we have obtained over angular derivatives as follows
\begin{eqnarray*}
\int_{s_1}^{s_2} \int_{{\cal H}_{s}} \left\vert \frac{m}{F} \hat{\eth} \hat{\psi}_1 \right\vert \vert \partial_R \hat{\psi}_0 \vert \frac{1}{\vert u \vert} \d u \d^2 \omega \d s & \lesssim & \int_{s_1}^{s_2} \frac{1}{\sqrt{s}}\int_{{\cal H}_{s}} \left\vert \frac{m}{F} \hat{\eth} \hat{\psi}_1 \right\vert \vert \partial_R \hat{\psi}_0 \vert \sqrt{\frac{R}{\vert u \vert}} \d u \d^2 \omega \d s \\
& \lesssim & \int_{s_1}^{s_2} \frac{1}{\sqrt{s}} \int_{{\cal H}_{s}} \left( \vert \hat{\eth} \hat{\psi}_1 \vert^2 + \frac{R}{\vert u \vert} \vert \partial_R \hat{\psi}_0 \vert^2 \right) \d u \d^2 \omega \d s \\
& \lesssim & \int_{s_1}^{s_2} \frac{1}{\sqrt{s}} \left( {\cal E}_{{\cal H}_{s}} (D_\omega \hat{\psi}) + {\cal E}_{{\cal H}_{s}} (D_R \hat{\psi}) \right) \d s \, ,
\end{eqnarray*}
which allows to apply a Gronwall inequality since $1/\sqrt{s}$ is integrable on $[0,1]$. The third term is the trickiest. We use the fact that the lowest eigenvalue of $\eth'$ on weighted scalar fields of weight $\{ 1 , 0 \}$ is positive (see \cite{PeRi84} section 4.15)~; this implies
\[ \int_{S^2} \vert \hat\psi_0 \vert^2 \d^2 \omega \lesssim \int_{S^2} \vert \hat\eth' \hat\psi_0 \vert^2 \d^2 \omega \]
uniformly on $\Omega^+_{u_0}$, whence using the same method as for the previous term,
\[ \int_{s_1}^{s_2} \int_{{\cal H}_{s}} \left\vert \frac{1- 5mR}{\sqrt{2F}} \hat{\psi}_0 \right\vert \vert \partial_R \hat{\psi}_0 \vert \frac{1}{\vert u \vert} \d u \d^2 \omega \d s \lesssim \int_{s_1}^{s_2} \frac{1}{\sqrt{s}} \int_{{\cal H}_{s}} ( \left\vert \hat{\eth}' \hat{\psi}_0 \right\vert^2 + \frac{R}{\vert u \vert} \vert \partial_R \hat{\psi}_0 \vert^2 ) \d u \d^2 \omega \d s \]
and since $\hat\eth' \hat{\psi}_0 = \hat{\mbox{\th}} \hat{\psi}_1$ which is controlled uniformly in $\Omega^+_{u_0}$ by $\vert \partial_R \hat{\psi}_1 \vert + \vert \hat{\psi}_1 \vert$, we can once again apply Gronwall's inequality. Successive applications of $\partial_R$ will produce error terms which can be controlled by similar techniques using angular derivatives and lower order norms. We get the following result~:
\begin{theorem} \label{DiracPeeling}
There exist positive constants $C_n$, $n\in \N$ such that for any smooth compactly supported data on ${\cal H}_1$, the associated solution $\hat{\psi}_A$ of Dirac's equation satisfies
\begin{gather*}
\sum_{p=0}^n \sum_{k=0}^p {\cal E}_{\scri^+_{u_0}}  (D^k_\omega D^{p-k}_R \hat\psi_A ) \leq C_n \sum_{p=0}^n \sum_{k=0}^p {\cal E}_{{\cal H}_1} (D^k_\omega D^{p-k}_R \hat\psi_A ) \, , \\
\sum_{p=0}^n \sum_{k=0}^p {\cal E}_{{\cal H}_1} (D^k_\omega D^{p-k}_R \hat\psi_A ) \leq C_n \sum_{p=0}^n \sum_{k=0}^p \left( {\cal E}_{\scri^+_{u_0}}  (D^k_\omega D^{p-k}_R \hat\psi_A ) + {\cal E}_{{\cal S}_{u_0}}  (D^k_\omega D^{p-k}_R \hat\psi_A ) \right) \, .
\end{gather*}
This extends to the spaces of initial data $\mathfrak{h}^n ({\cal H}_1 )$ obtained by completion of ${\cal C}^\infty_0 ({\cal H}_1 )$ in the norms
\[ \Vert \hat\psi_A \Vert_{\mathfrak{h}^n ({\cal H}_1 )} = \left( \sum_{p=0}^n \sum_{k=0}^p {\cal E}_{{\cal H}_1} (D^k_\omega D^{p-k}_R \hat\psi_A ) \right)^{1/2} \, . \]
\end{theorem}

\subsection{Peeling for Maxwell}

Contrary to the case of the Dirac equation, we already have a problem with the basic energy estimate since the energy current $V$ does not satisfy an exact conservation law. However, the error term in the approximate conservation law satisfied by $V$ is easily controlled by the energy density on the hypersurfaces ${\cal H}_{s}$. This gives us the following result~:
\begin{proposition} \label{BasicEstimateMax}
There exists a positive positive constant $C$, such that for any smooth data on ${\cal H}_1$ satisfying the constraints and supported away from $i^0$, the associated solution $\hat{\phi}_{AB}$ of Maxwell's equations satisfies for any $0\leq s<1$
\begin{gather*}
{\cal E}_{{\cal H}_s}  (\hat\phi_{AB} ) \leq C {\cal E}_{{\cal H}^1} (\hat\phi_{AB} ) \, , \\
{\cal E}_{{\cal H}_1} ( \hat\phi_{AB} ) \leq C \left( {\cal E}_{{\cal H}_s}  (\hat\phi_{AB} ) + {\cal E}_{{\cal S}^{s,1}_{u_0}}  (\hat\phi_{AB} ) \right) \, .
\end{gather*}
\end{proposition}
We now obtain similar estimates for successive derivatives of the Maxwell field, starting with the derivative with respect to $R$. We multiply by $\sqrt{2F}$ the first and third equations of \eqref{GHPMaxwell} and commute $\partial_R$ into the system. We obtain
\[
\left\{ \begin{array}{l}
{ \left( 2\partial_u + R^2 F \partial_R - (5mR^2 -2R ) \right) \partial_R \hat{\phi}_0 - \sqrt{2F} \, \hat\eth \partial_R \hat{\phi}_1 } \\
{ \hspace{1in} = -2R(1-3mR) \partial_R \hat\phi_0 -2 (1-5mR) \hat\phi_0 - \sqrt{\frac{2}{F}} m \hat\eth \hat\phi_1  \, , } \\ \\
{ -\sqrt{\frac{F}{2}} \partial_R^2 \hat{\phi}_1 - \hat{\eth}' \partial_R \hat{\phi}_0 = -\frac{m}{\sqrt{2F}} \partial_R \hat\phi_1 \, ,} \\ \\
{ (2 \partial_u + R^2F\partial_R ) \partial_R \hat{\phi}_1 - \sqrt{2F} \hat{\eth} \partial_R \hat{\phi}_2 = -2R(1-3mR)  \partial_R \hat\phi_1 - \sqrt{\frac{2}{F}} m \eth \hat\phi_2 \, , } \\ \\
{ -\sqrt{\frac{F}{2}} \partial_R^2 \hat{\phi}_2 + \frac{m}{\sqrt{2F}} \partial_R \hat\phi_2 - \hat{\eth}' \partial_R \hat{\phi}_1 = -\frac{m}{\sqrt{2F}} \partial_R \hat\phi_2 - \frac{m^2}{(2F)^{3/2}} \hat\phi_2 \, , }   \end{array} \right.
\]
which we rewrite as
\[
\left\{ \begin{array}{l}
{ \hat\thorn ' \partial_R \hat{\phi}_0 - \hat\eth \partial_R \hat{\phi}_1 = -\sqrt{\frac{2}{F}} R(1-3mR) \partial_R \hat\phi_0 -\sqrt{\frac{2}{F}} (1-5mR) \hat\phi_0 - \frac{m}{F} \hat\eth \hat\phi_1  \, , } \\ \\
{ \hat\thorn \partial_R \hat{\phi}_1 - \hat{\eth}' \partial_R \hat{\phi}_0 = -\frac{m}{\sqrt{2F}} \partial_R \hat\phi_1 \, ,} \\ \\
{ \hat\thorn' \partial_R \hat{\phi}_1 - \hat{\eth} \partial_R \hat{\phi}_2 = -\sqrt{2}{F} R(1-3mR)  \partial_R \hat\phi_1 - \frac{m}{F} \eth \hat\phi_2 \, , } \\ \\
{ \hat\thorn \partial_R \hat{\phi}_2 - \hat{\eth}' \partial_R \hat{\phi}_1 = -\frac{m}{\sqrt{2F}} \partial_R \hat\phi_2 - \frac{m^2}{(2F)^{3/2}} \hat\phi_2 \, , }   \end{array} \right.
\]
Putting
\[ D_R \hat\phi_{AB} := \partial_R \hat\phi_0 \hat\iota_A \hat\iota_B - \partial_R \hat\phi_1 (\hat{o}_A \hat{\iota}_B + \hat\iota_A \hat{o}_B ) + \partial_R \hat\phi_2 \hat{o}_A \hat{o}_B \, ,\]
the above system is the perturbed Maxwell equation for $D_R \hat\phi_{AB}$~:
\begin{eqnarray*}
\nabla^{AA'} D_R \hat\phi_{AB} &=& - \left( \sqrt{2} FR (1-3mR) \partial_R \hat\phi_1 + \frac{m}{F} \hat\eth \hat\phi_2 \right) \bar{\hat{o}}^{A'} \hat{o}_B \\
&& + \left( \sqrt{\frac{2}{F}} R (1-3mR) \partial_R \hat\phi_0 + \sqrt{\frac{2}{F}} (1-5mR) \hat\phi_0 + \frac{m}{F} \hat\eth \hat\phi_1 \right) \bar{\hat{o}}^{A'} \hat\iota_B \\
&& - \left( \frac{m}{\sqrt{2F}} \partial_R \hat\phi_2 + \frac{m^2}{(2F)^{3/2}} \hat{\phi}_2 \right) \hat\iota^{A'} \hat{o}_B + \frac{m}{\sqrt{2F}} \partial_R \hat{\phi}_1 \hat\iota^{A'} \hat\iota_B \, .
\end{eqnarray*}
The associated approximate conservation law is therefore
\begin{gather*}
\nabla^{AA'} \left( T^{BB'} (D_R \hat\phi_{AB}) (D_R \bar{\hat\phi}_{A'B'}) \right) = \nabla^{(a} T^{b)} (D_R \hat\phi_{AB}) (D_R \bar{\hat\phi}_{A'B'}) \\
+ 2 \Re \left\{ T^{BB'} \left[ - \left( \sqrt{2} FR (1-3mR) \partial_R \hat\phi_1 + \frac{m}{F} \hat\eth \hat\phi_2 \right) \bar{\hat{o}}^{A'} \hat{o}_B \right. \right.\\
+ \left( \sqrt{\frac{2}{F}} R (1-3mR) \partial_R \hat\phi_0 + \sqrt{\frac{2}{F}} (1-5mR) \hat\phi_0 + \frac{m}{F} \hat\eth \hat\phi_1 \right) \bar{\hat{o}}^{A'} \hat\iota_B \\
\left. \left. - \left( \frac{m}{\sqrt{2F}} \partial_R \hat\phi_2 + \frac{m^2}{(2F)^{3/2}} \hat{\phi}_2 \right) \hat\iota^{A'} \hat{o}_B + \frac{m}{\sqrt{2F}} \partial_R \hat{\phi}_1 \hat\iota^{A'} \hat\iota_B \right] D_R \bar{\hat{\phi}}_{A'B'} \right\} \, ,
\end{gather*}
which in full details, using the decomposition \eqref{MorawetzTetrad} of $T^a$ on the rescaled tetrad, reads
\begin{eqnarray}
&&\nabla^{AA'} \left( T^{BB'} (D_R \hat\phi_{AB}) (D_R \bar{\hat\phi}_{A'B'}) \right) = 8mR^2 F^{-1} (3+uR) \vert \partial_R \hat\phi_0 \vert^2 \nonumber \\
&& \hspace{1in} - \frac{4}{F} \left( 2(1+uR) + \frac{(uR)^2F}{2} \right) (1-3mR) R \vert \partial_R \hat\phi_0 \vert^2 \nonumber \\
&& \hspace{1in} - \left( 2 F^{3/2} (1-3mR) \vert Ru\vert \vert u \vert + \frac{m}{F} \left( 2(1+uR) + \frac{(uR)^2F}{2} \right) \right) \vert \partial_R \hat\phi_1 \vert^2 \nonumber \\
&& \hspace{1in} - mu^2 \vert \partial_R \phi_2 \vert^2 \nonumber \\
&& \hspace{1in} - 2 \Re \left[ \frac{m}{\sqrt{2F}} u^2 \hat\eth \hat\phi_2 \partial_R \bar{\hat{\phi}}_1 + \frac{m^2}{4F} u^2 \hat\phi_2 \partial_R \bar{\hat\phi}_2 \right. \nonumber \\
&& \hspace{1.5in} + \frac{2}{F} (1-5mR) \left( 2(1+uR) + \frac{(uR)^2F}{2} \right) \hat\phi_0 \partial_R \bar{\hat\phi}_0 \nonumber \\
&& \hspace{1.5in} \left. + \frac{\sqrt{2} m}{F^{3/2}} \left( 2(1+uR) + \frac{(uR)^2F}{2} \right) \hat\eth \hat\phi_1 \partial_R \bar{\hat{\phi}}_0 \right] \, . \label{MaxDRConsLaw}
\end{eqnarray}
As was the case for the Weyl equation, it is necessary to control the angular derivatives first before gaining a control on the derivative with respect to $R$. However, the weights of $\hat\phi_0$, $\hat\phi_1$ and $\hat\phi_2$ are $\{ 2,0 \}$, $\{ 0,0 \}$ and $\{ -2 , 0 \}$ and for an anti-self-dual field $\hat\phi_{A'B'}$ we would get the weights $\{ 0,2 \}$, $\{ 0,0 \}$ and $\{ 0,-2 \}$~; since the weights of $\hat\eth$ and $\hat\eth '$ are $\{ 1, -1 \}$ and $\{ -1 , 1 \}$, by commuting these operators into the Maxwell system we do not recover any of the adequate weights of the components. So we cannot hope to commute a well-chosen arrangement of $\hat\eth$ and $\hat\eth '$ into the (anti-self-dual) Maxwell system and obtain another (self-dual) Maxwell system. But the Schwarzschild metric is spherically symmetric, we have a $3$-dimensional space of Killing vector fields tangent to the sphere which generate all rotations. We choose a basis of this space, denoted $X$, $Y$, $Z$, such that $-X^2 - Y^2 - Z^2$ is controlled below and above by the positive Laplacian on $S^2$. Since $X$, $Y$ and $Z$ will commute with Maxwell's equations, this gives us a control analogous to proposition \ref{BasicEstimateMax} over angular derivatives of any order.

Integrating the error term of \eqref{MaxDRConsLaw} over $\Omega^+_{u_0} \cap \{ \tau \leq s \leq1 \}$, for $0<\tau \leq 1$, and splitting the result as an integral in $s$ of integrals over the hypersurfaces ${\cal H}_s$, we gain a factor $1/u$ in the $3$-volume measure. With this, the integrals of all terms on ${\cal H}_s$ are controlled for each $s$ by the sum
\[ {\cal E}_{{\cal H}_s} (\hat\phi_{AB} ) + {\cal E}_{{\cal H}_s} (D_R \hat\phi_{AB} ) + {\cal E}_{{\cal H}_s} (D_X \hat\phi_{AB} ) + {\cal E}_{{\cal H}_s} (D_Y \hat\phi_{AB} ) + {\cal E}_{{\cal H}_s} (D_Z \hat\phi_{AB} )  \, ,\]
except for the last two terms. The treatment for both terms is similar to the Dirac case. For the first, we use the fact that the $L^2$ norm of $\hat\phi_0$ on the $2$-sphere is (uniformly in $\Omega_{u_0}^+$) controlled\footnote{The first eigenvalue of $\hat\eth'$ on weighted scalars of weight $\{ 2 , 0 \}$ is again positive, see \cite{PeRi84} section 4.15.} by that of $\hat\eth' \hat\phi_0$ and then the second equation of the Maxwell system giving us the equality of $\hat\eth' \hat\phi_0$ and $\hat\thorn \hat\phi_1$. So we obtain for $0<s \leq 1$
\begin{gather*}
2\int_{{\cal H}_s} \left\vert \frac{2}{F} (1-5mR) \left( 2(1+uR) + \frac{(uR)^2F}{2} \right) \hat\phi_0 \partial_R \bar{\hat\phi}_0 \right\vert \frac{1}{\vert u \vert } \d u \d^2 \omega \\
\hspace{0.5in} \lesssim \frac{1}{\sqrt{s}} \int_{{\cal H}_s} \vert \hat\phi_0  \partial_R \bar{\hat\phi}_0 \vert \sqrt\frac{R}{\vert u \vert } \d u \d^2 \omega \\
\hspace{0.5in} \lesssim \frac{1}{\sqrt{s}} \left( \int_{{\cal H}_s} \vert \hat\eth '\hat\phi_0 \vert^2 \d u \d^2 \omega + \int_{{\cal H}_s} \vert \partial_R \bar{\hat\phi}_0 \vert^2 \frac{R}{\vert u \vert } \d u \d^2 \omega \right) \\
\hspace{0.5in} \lesssim \frac{1}{\sqrt{s}} \left( \int_{{\cal H}_s} \left( \vert \hat\phi_1 \vert^2 + \vert \partial_R \hat\phi_1 \vert^2 \right) \d u \d^2 \omega + \int_{{\cal H}_s} \vert \partial_R \bar{\hat\phi}_0 \vert^2 \frac{R}{\vert u \vert } \d u \d^2 \omega \right) \\
\hspace{0.5in} \lesssim \frac{1}{\sqrt{s}} \left( {\cal E}_{{\cal H}_s} (\hat\phi_{AB} ) + {\cal E}_{{\cal H}_s} (D_R \hat\phi_{AB} ) \right)\, .
\end{gather*}
The second is simpler to deal with~:
\begin{gather*}
2\int_{{\cal H}^s} \left\vert  \frac{\sqrt{2} m}{F^{3/2}} \left( 2(1+uR) + \frac{(uR)^2F}{2} \right) \hat\eth \hat\phi_1 \partial_R \bar{\hat{\phi}}_0 \right\vert \frac{1}{\vert u \vert } \d u \d^2 \omega \\
\hspace{0.5in} \lesssim \frac{1}{\sqrt{s}} \int_{{\cal H}^s} \vert \hat\eth \hat\phi_1  \partial_R \bar{\hat\phi}_0 \vert \sqrt\frac{R}{\vert u \vert } \d u \d^2 \omega \\
\hspace{0.5in} \lesssim \frac{1}{\sqrt{s}} \left( {\cal E}_{{\cal H}^s} (D_X \hat\phi_{AB} ) + {\cal E}_{{\cal H}^s} (D_Y \hat\phi_{AB} ) + {\cal E}_{{\cal H}^s} (D_Z \hat\phi_{AB} ) + {\cal E}_{{\cal H}^s} (D_R \hat\phi_{AB} ) \right)\, .
\end{gather*}
\begin{remark}
Note that since $\hat\phi_1$ is the component of $\hat\phi_{AB}$ whose weight is $\{ 0,0 \}$, $\hat\eth \hat\phi_1$ is merely $\hat\delta \hat\phi_1 = \nabla_{\hat{m}} \hat\phi_1$, it involves no spin-coefficient. And since $m$ is a bounded vector field on $S^2$, the $L^2$ norm of $\hat\eth \hat\phi_1$ on $S^2$ is controlled (uniformly on $\Omega^+_{u_0}$) by the sum of the $L^2$ norms of $D_X \hat\phi_1$, $D_Y \hat\phi_1$ and $D_Z \hat\phi_1$.
\end{remark}
The successive derivatives with respect to $R$ will be controlled in a similar way, controlling first the angular derivatives. This gives the theorem~:
\begin{theorem} \label{PeelingMaxwell}
There exist positive constants $C_n$, $n\in \N$ such that for any smooth data on ${\cal H}_1$ compactly supported, the associated solution $\hat{\phi}_{AB}$ of Maxwell's equations satisfies
\begin{gather*}
\sum_{k_1+k_2+k_3+k_4\leq n} {\cal E}_{\scri^+_{u_0}}  (D^{k_1}_X D^{k_2}_Y D^{k_3}_Z D^{k_4}_R \hat\phi_{AB} ) \leq C_n \sum_{k_1+k_2+k_3+k_4\leq n} {\cal E}_{{\cal H}_1} (D^{k_1}_X D^{k_2}_Y D^{k_3}_Z D^{k_4}_R \hat\phi_{AB} ) \, , \\
\sum_{k_1+k_2+k_3+k_4\leq n} {\cal E}_{{\cal H}_1} (D^{k_1}_X D^{k_2}_Y D^{k_3}_Z D^{k_4}_R \hat\phi_{AB} ) \leq C_n \sum_{k_1+k_2+k_3+k_4\leq n} \left( {\cal E}_{\scri^+_{u_0}}  (D^{k_1}_X D^{k_2}_Y D^{k_3}_Z D^{k_4}_R \hat\phi_{AB} ) \right. \\
\hspace{4in} \left.+ {\cal E}_{{\cal S}_{u_0}}  (D^{k_1}_X D^{k_2}_Y D^{k_3}_Z D^{k_4}_R \hat\phi_{AB} ) \right) \, .
\end{gather*}
This extends to the spaces of initial data $\mathfrak{h}^n ({\cal H}_1 )$ obtained by completion of ${\cal C}^\infty_0 ({\cal H}_1 )$ in the norms
\[ \Vert \hat\phi_{AB} \Vert_{\mathfrak{h}^n ({\cal H}_1 )} = \left( \sum_{k_1+k_2+k_3+k_4\leq n} {\cal E}_{\scri^+_{u_0}}  (D^{k_1}_X D^{k_2}_Y D^{k_3}_Z D^{k_4}_R \hat\phi_{AB} ) \right)^{1/2} \, . \]
\end{theorem}

\section{Interpretation} \label{Interpretation}

In this section, we check that for a given order of transverse regularity at $\scri^+$, our classes of data ensuring that the rescaled solution has at least this regularity are not smaller than they are in the flat case when the full embedding in the Einstein cylinder is used. Of course, since we have only worked in a neighbourhood of $i^0$ in Schwarzschild, what we really mean by this is a comparison of the asymptotic constraints on the fall-off of initial data. In the case of Dirac and Maxwell fields, this comparison is made easier than for the wave equation (see \cite{MaNi2009} for the analogous interpretation for the wave equation) because the energy on a spacelike slice is conformally invariant.

Let us consider a $4$-dimensional globally hyperbolic (and therefore admitting a spin-structure) spacetime $({\cal M} , g)$, $\Sigma$ a spacelike hypersurface, $\nu^a$ its future-oriented unit normal vector field, $\psi_A$ a solution to \eqref{WeylEq} and $\Phi_{AB}$ a solution to \eqref{MaxwellEq}. Let $\Omega$ be a smooth positive function on $\cal M$, and put
\[ \hat{g} := \Omega^2 g \, ,~ \hat{\psi}_A := \Omega^{-1} \psi_A \, ,~ \hat{\phi}_{AB} := \Omega^{-1} \phi_{AB} \, .\]
The unit normal to $\Sigma$ for $\hat{g}$ is now
\[ \hat{\nu}^a = \Omega^{-1} \nu^a \]
and if we denote by $\mu$ (resp. $\hat{\mu}$) the measure induced on $\Sigma$ by $g$ (resp. $\hat{g}$), then
\[ \hat{\mu} = \Omega^3 \mu \, .\]
The energy of the rescaled Weyl field on $\Sigma$ is given by (in this section, we shall denote by $\hat{\cal E}$ the energies for the rescaled metric and $\cal E$ the energies for the unrescaled metric)
\[ \hat{\cal E}_\Sigma (\hat\psi ) = \int_\Sigma \hat{\nu}^a \hat{\psi}_A \bar{\hat{\psi}}_{A'} \d \hat{\mu} = \int_\Sigma {\nu}^a {\psi}_A \bar{\psi}_{A'} \d {\mu} = {\cal E}_\Sigma (\psi ) \, .\]
For the Maxwell field, we need a choice of observer (or merely of timelike vector field) $t^a$ in the neighbourhood of $\Sigma$ to define the energy and then
\[ \hat{\cal E}_\Sigma (\hat\phi ) = \int_\Sigma \hat{\nu}^a t^b \hat{\phi}_{AB} \bar{\hat{\phi}}_{A'B'} \d \hat{\mu} = \int_\Sigma {\nu}^a t^b {\phi}_{AB} \bar{\phi}_{A'B'} \d {\mu} = {\cal E}_\Sigma (\phi ) \, .\]
Note that the observer is not rescaled.

\subsection{Dirac fields}

We now compare the classes of data for Dirac fields in the flat case and in the Schwarzschild spacetime. In the flat case ($m=0$), the conformal embedding of Minkowski spacetime into the Einstein cylinder is realized using the conformal factor
\[ \Omega = \frac{2}{\sqrt{1+(t+r)^2} \sqrt{1+(t-r)^2}} \]
and the vector field used for increasing the regularity in the energy estimates is the time translation along the Einstein cylinder
\[ \frac{\partial}{\partial \tau} = \frac{1}{2} \left( (1+t^2+r^2) \frac{\partial}{\partial t} + 2tr \frac{\partial}{\partial r} \right) \, . \]
This is a Killing vector field on the Einstein cylinder, i.e. a conformal Killing vector field of Minkowski spacetime. Let $\psi_A$ be a Weyl field on Minkowski spacetime, denoting by $\Sigma$ the $t=0$ hypersurface and choosing the spin-frame given by the choice of Newman-Penrose tetrad
\begin{equation} \label{FlatNP}
l = \frac{1}{\sqrt{2}} (\partial_t + \partial_r ) \, ,~ n = \frac{1}{\sqrt{2}} (\partial_t + \partial_r ) \, ,~ m=\frac{1}{\sqrt{2}} (\partial_\theta +\frac{i}{\sin \theta} \partial_\varphi ) \, ,
\end{equation}
the future-oriented unit (for the unrescaled metric) normal to $\Sigma$ is $\partial_t = \frac{1}{\sqrt{2}} (l+n )$ and we have
\[ \hat{\cal E}_\Sigma (\hat\psi ) = {\cal E}_\Sigma (\psi ) = \int_\Sigma \nu^a \psi_A \bar\psi_{A'} \d^3 x = \frac{1}{\sqrt{2}} \int_\Sigma (\vert \psi_0 \vert^2 + \vert \psi_1 \vert^2 ) \d^3 x \, .\]
On the Einstein cylinder, we have the following energy equality
\[ \hat{\cal E}_\Sigma (\hat\psi ) = \hat{\cal E}_{\scri^+} (\hat\psi ) \]
and commuting $\partial_\tau^k$ into the equation we get
\[ \hat{\cal E}_\Sigma (\partial_\tau^k \hat\psi ) = \hat{\cal E}_{\scri^+} (\partial_\tau^k \hat\psi ) \, .\]
The right-hand side is a measure of transverse regularity at $\scri^+$.

We work out the constraint on initial data corresponding to the first level of regularity. Using the expression of $\partial_\tau$ in terms of $(t,r)$ variables at $t=0$, the left-hand side for $k=1$ can be rewritten as
\[ \hat{\cal E}_\Sigma (\frac{1+r^2}{2} \partial_t \hat\psi ) ={\cal E}_\Sigma (\frac{1+r^2}{2} \partial_t \psi ) \, .\]
Using the ellipticity of the spacelike part of the Dirac equation (i.e. using the Bochner-Lichnerowicz-Weitzenböck formula on $\R^3$), this corresponds, modulo lower order terms, to an $L^2$ control over
\[ (1+r^2) \partial_r \psi \mbox{ and } \frac{1+r^2}{r} \nabla_{S^2} \psi \]
independently. This is what defines the scale of weighted Sobolev spaces obtained on $\Sigma$ by requiring the finiteness of the energies of $\partial_\tau^k \hat\psi$.

Note that this is not the original conformal peeling construction by Penrose, but it is the closest equivalent in terms of Sobolev spaces (see \cite{MaNi2009} for a more detailed presentation for the wave equation).

In the Schwarzschild case, the energy equality is now
\[ \hat{\cal E}_{{\cal H}^1} (\hat\psi ) = \hat{\cal E}_{{\cal S}_{u_0}} (\hat\psi ) + \hat{\cal E}_{\scri^+_{u_0}} (\hat\psi ) \]
and to raise the regularity up to order $k$, we use all combinations up to order $k$ of $\partial_R$ and $D_{\omega}$. At first order, the quantity defining the space of data on ${\cal H}^1$ is
\begin{equation} \label{FirstOrderNorm}
\hat{\cal E}_{{\cal H}^1} (\hat\psi ) + \hat{\cal E}_{{\cal H}^1} (\partial_R \hat\psi ) + \hat{\cal E}_{{\cal H}^1} (D_\omega \hat\psi ) \, ,
\end{equation}
where $D_\omega$ can be replaced by $\nabla_{S^2}$, the former acting on the components, the latter on the full spinor. First of all, using the conformal invariance of the energy, for the unrescaled spin-frame associated with the tetrad \eqref{OriginalTetrad},
\[ \hat{\cal E}_{{\cal H}^1} (\hat\psi ) = \int_{{\cal H}^1} (\vert \psi_0 \vert^2 + \vert \psi_1 \vert^2) F^{1/2} r^2 \sin \theta \d r \d \theta \d \varphi \, ,\]
which gives a control equivalent to the flat $L^2$ norm. Now up to lower order terms, using the conformal invariance of the energy, \eqref{FirstOrderNorm} is equivalent to the same expression with unrescaled energies for the unrescaled field $\psi$. The last term in \eqref{FirstOrderNorm} bounds the $L^2$ norm of $\nabla_{S^2} \psi$ with no weight, which is weaker than in the flat case, but the second term will impose a constraint similar to the flat case, though in a more mixed manner. Indeed, using the equation satisfied by $\psi$ in components (see for example \cite{Ni1997})
\[ \partial_t \psi + F\left( \begin{array}{cc} {-1} & 0 \\ 0 & 1 \end{array} \right) (\partial_r + \frac{1}{r} + \frac{F'}{4F} ) \psi + \frac{F^{1/2}}{r} D_\omega \psi =0 \, ,\]
we find that
\begin{eqnarray*}
\partial_R \psi &=& \frac{r^2}{F} (\partial_t + \partial_{r_*} ) \psi = \frac{2r^2}{F} \left( \begin{array}{c} {\partial_{r_*} \psi_0 } \\ 0 \end{array} \right) - r F^{-1/2} D_\omega \psi + \mbox{ lower order terms,} \\
&=& \left( \begin{array}{c} {2r^2 \partial_{r} \psi_0 } \\ 0 \end{array} \right) - r F^{-1/2} D_\omega \psi + \mbox{ lower order terms.}
\end{eqnarray*}
We see that in terms of weights, this is similar to the flat case. The gain is that we only control $r^2 \partial_r$ applied to $\psi_0$ and not to $\psi_1$. This is not surprising since $\psi_0$ is the component of $\psi$ that propagates dominantly to the left (towards the black hole) and which, were its fall-off at infinity not strong enough, might therefore propagate singularities along $\scri^+$ from $i^0$. Also the weight in front of the angular derivatives in the expression of $\partial_R \psi$ is the same as in the flat case, but the term is not controlled on its own, only in combination with $r^2 \partial_{r} \psi_0$. This is what defines the scale of weighted Sobolev spaces obtained from our energy estimates in the Schwarzschild case.

The conclusion is that the constraints for peeling in the Schwarzschild case are weaker than the ones obtained in Minkowski using the full conformal embedding in the Einstein cylinder. In other words, peeling in Schwarzschild at any order is valid for a class of data slightly larger than the usual class in the flat case.
\begin{remark}
This does not mean that the asymptotic structure of Schwarzschild allows peeling for more general data than in Minkowski. This only shows that our definition of peeling is more general than the one usually considered. The uniformity of our norms in the mass in any compact interval $[0,M]$ shows that for our definition, the classes in Minkowski and Schwarzschild are the same.
\end{remark}

\subsection{Maxwell fields}

In the Maxwell case, the essential ingredients are the same as in the Dirac case~: conformal invariance of the norm, same vector fields used to raise the regularity and controlling better the part of the data propagating to the left, plus one more ingredient, the Morawetz vector field. What is not clear for Maxwell fields is that the basic energies should give equivalent controls. The reason why this is true is that in flat spacetime, the Morawetz vector field and the time translation along the Einstein cylinder differ by a constant multiple of $\partial_t$ (see \cite{MaNi2009}) which gives a weaker norm than the two others, plus the local equivalence of the norms in the mass $m$. Let us describe this more explicitely.

On the Einstein cylinder, we use the vector field $\partial_\tau$ both for defining the conserved current for the Maxwell field and for raising the regularity in the energy equalities. Using the conformal invariance of the energy and the expression of $\partial_\tau$, decomposing the unrescaled field in the spin-frame associated with \eqref{FlatNP}, we get
\[ \hat{\cal E}_\Sigma (\hat\phi ) = {\cal E}_\Sigma (\phi ) = \int_\Sigma ( \vert \phi_0 \vert^2 + 2 \vert \phi_1 \vert^2 + \vert \phi_2 \vert^2 )\frac{1+r^2}{4} \d^3 x \, .\]
In the Schwarzschild case, we define the energy using the Morawetz vector field given in terms of the vectors of the rescaled tetrad by \eqref{MorawetzTetrad}. The initial energy therefore is given by (using the fact that the future unit normal to the hypersurface $\{ t=0 \}$ for the metric $g$ is $\frac{1}{\sqrt{2}} (l+n)$ and also $u=-r_*$ on ${\cal H}^1$)
\begin{eqnarray*}
\hat{\cal E}_{{\cal H}^1} (\hat\phi ) = {\cal E}_{{\cal H}^1} (\phi ) &=& \int_{{\cal H}^1} \left( \frac{1}{\sqrt{F}} ( 2(1-r_*R) + \frac{(r_*R)^2F}{2} ) r^2 \vert \phi_0 \vert^2 \right. \\
&& + ( \frac{1}{\sqrt{F}} (2(1-r_*R) + \frac{(r_*R)^2F}{2} ) r^2 + \frac{r_*^2 \sqrt{F}}{2} )\vert \phi_1 \vert^2 \\
&& \left. + \frac{\sqrt{F} r_*^2}{2} \vert \phi_2 \vert^2 \right) \sqrt{F} r^2 \d r_* \d \omega \, ,
\end{eqnarray*} 
which in $\Omega_{u_0}^+$ is equivalent to
\[ \int_{{\cal H}^1} \left( \vert \phi_0 \vert^2 + \vert \phi_1 \vert^2 + \vert \phi_2 \vert^2 \right)  r^4 \d r \d \omega \]
and therefore gives the same control as the flat norm involving $\partial_\tau$ instead of the Morawetz vector field. When raising the regularity in the energy estimates, the principles are the same as for Dirac. On Minkowski spacetime, a $\partial_t$ applied to $\phi$ gives a control on the full gradient on $\R^3$ because Maxwell equations entail that the d'Alembertian of the field vanishes and this ensures the ellipticity of the spacelike part. So the flat spacetime energy at the first level of regularity satisfies
\[ {\cal E}_\Sigma ( \phi ) + {\cal E}_\Sigma (\frac{1+r^2}{2} \partial_t \phi ) \simeq \int_\Sigma \left( \vert \phi \vert^2 + (1+r^2) \vert \nabla_{\R^3}  \phi \vert^2 \right) (1+r^2) \d^3 x \, .\]
In the Schwarzschild case, the first level of regularity is controlled by the energy
\[ \hat{\cal E}_{{\cal H}^1} (\hat\phi ) + \hat{\cal E}_{{\cal H}^1} (\partial_R \hat\phi ) + \hat{\cal E}_{{\cal H}^1} (\nabla_{S^2} \hat\phi ) \, , \]
which is equivalent to
\[ {\cal E}_{{\cal H}^1} (\phi ) + {\cal E}_{{\cal H}^1} (\partial_R \phi ) + {\cal E}_{{\cal H}^1} ( \nabla_{S^2} \phi ) \, . \]
Similarly to the Dirac case, we can evaluate $\partial_R \phi$ using
\[ \partial_R = \frac{r^2}{F} (\partial_t + \partial_{r_*} ) \]
and Maxwell's equations. We have~:
\begin{eqnarray*}
\partial_R \phi_0 &=& \frac{r^2}{F} ( \sqrt{2F} m^a \partial_a \phi_1 + 2 \partial_{r_*} \phi_0 )+\mbox{ lower order terms,} \\
\partial_R \phi_1 &=& \sqrt{\frac{2}{F}} r^2 \bar{m}^a \partial_a \phi_0 +\mbox{ lower order terms,} \\
\partial_R \phi_2 &=& \sqrt{\frac{2}{F}} r^2 \bar{m}^a \partial_a \phi_1 +\mbox{ lower order terms.}
\end{eqnarray*}
So we observe exactly the same phenomenon as for Dirac, a similar control as in the flat case, except that the control on radial derivatives is weaker~: only the component propagating to the left is explicitely controlled on the initial hypersurface.

\subsection{A remark on the constraints}

This paper is concerned about the asymptotic assumptions on initial data ensuring peeling at a certain order and on comparing these asumptions in Schwarzschild and in flat spacetime. There is one aspect, for Maxwell fields, which may affect the classes of data leading to a peeling at a certain order in different ways in Schwarzschild and Minkowski~: the constraints. We do not address this question in detail here but we simply make a remark on the integrability of the constraints from infinity in a reasonably large class of functions and the compatibility between the constraints and our energy norms. Let us consider the Maxwell system in the unrescaled Schwarzschild spacetime using the tetrad \eqref{OriginalTetrad}. The energy at $t=0$ associated with the Morawetz vector field in the $r_* > -u_0$ region is equivalent to
\[ \int_{r_* >-u_0} (\vert \phi_0 \vert^2 + \vert \phi_1 \vert^2 + \vert \phi_2 \vert^2 ) r^4 \d r_* \d^2 \omega \, .\]
Taking the difference of the second and third equations of the system, we get
\[ F \left( \partial_r + \frac{2}{r} \right) \phi_1 = \frac{1}{r} \left( \eth \phi_2 - \eth' \phi_0 \right) \, ;\]
rescaling the field components by $r^2$,
\[ \tilde{\phi}_i := r^2 \phi_i \, ,~i=0,1,2,\]
this reads
\begin{equation} \label{RescConstraints}
F \partial_r \tilde{\phi}_1 = \frac{1}{r} \left( \eth \tilde\phi_2 - \eth' \tilde\phi_0 \right) \, .
\end{equation}
Decomposing the field into spin-weighted spherical harmonics, the operators $\eth$ and $\eth'$ will turn into multiplication operators by constant factors. If we consider a simplified situation where the behaviour of the field at infinity is in powers of $r^{-1}$, the finiteness of the Morawetz energy implies that the rescaled field components fall off at least like $r^{-1}$. Assuming such a behaviour for $\tilde\phi_0$ and $\tilde\phi_2$ and integrating \eqref{RescConstraints} from infinity starting from the value zero, we recover the same fall-off for $\tilde\phi_1$.

So heuristically, the constraints are compatible with the Morawetz energy in both the Schwarz\-schild and Minkowski spacetimes. This heuristic argument does not address the question of polyhomogeneous solutions of the constraints. Our sole purpose here is to show that in Schwarzschild and Minkowski, there will be large classes of solutions of the constraints compatible with our function spaces and that the constraints will not introduce in the Schwarzschild case any additional restriction compared to the flat case.

\appendix

\section{Covariant derivative approach} \label{App1}

\subsection{Curvature spinors}

We calculate the curvature spinors for the rescaled metric. Recall that given a spacetime $({\cal M},g)$ with a spin structure and equipped with the Levi-Civitta connection, the Riemann tensor $R_{abcd}$ can be decomposed as follows (see \cite{PeRi84})~:
\[ R_{abcd} = X_{ABCD} \, \varepsilon_{A'B'} \varepsilon_{C'D'} + \Phi_{ABC'D'} \, \varepsilon_{A'B'} \varepsilon_{CD} + \bar{\Phi}_{A'B'CD} \, \varepsilon_{AB} \varepsilon_{C'D'} + \bar{X}_{A'B'C'D'} \, \varepsilon_{AB} \varepsilon_{CD} \, ,\]
where $X_{ABCD}$ is a complete contraction of the Riemann tensor in its primed spinor indices
\[ X_{ABCD} = \frac{1}{4} R_{abcd} {\varepsilon}^{A'B'} {\varepsilon}^{C'D'} \]
and ${\Phi}_{ab} = {\Phi}_{(ab)}$ is the trace-free part of the Ricci tensor multiplied by $-1/2$~:
\[ 2{\Phi}_{ab} = 6 {\Lambda} {g}_{ab} - {R}_{ab} \, ,~ {\Lambda} = \frac{1}{24} \mathrm{Scal}_{g} \, . \]
It is convenient when dealing with conformal rescalings to use instead of $\Phi_{ab}$ the curvature $2$-form
\[ P_{ab} = \Phi_{ab} - \Lambda g_{ab} \]
because of its simpler transformation law. Also, it is usual to isolate the totally symmetric part $\Psi_{ABCD} $ of $X_{ABCD}$, referred to as the Weyl spinor, which describes the conformally invariant part of the curvature~:
\[ X_{ABCD} = \Psi_{ABCD} + \Lambda \left( \varepsilon_{AC} \varepsilon_{BD} + \varepsilon_{AD} \varepsilon_{BC} \right) \, ,~ \Psi_{ABCD} = X_{(ABCD)} \, . \]
The scalar $\Lambda$ and the curvature spinors $P$ and $\Psi$ have simple rules of transformation under a conformal rescaling $\hat{g} = \Omega^{2} g$, given by (see \cite{PeRi84} p. 120-123)~:
\begin{gather*}
\hat{\Psi}_{ABCD} = \Psi_{ABCD} \, , \\
\hat{\Lambda} = \Omega^{-2} \Lambda + \frac{1}{4} \Omega^{-3} \square \Omega \, , ~\square = \nabla^a \nabla_a \, , \\
\hat{P}_{ab} = P_{ab} - \nabla_b \Upsilon_a  + \Upsilon_{AB'} \Upsilon_{BA'} \, ,~ \mbox{with } \Upsilon_a = \Omega^{-1} \nabla_a \Omega = \nabla_a \log \Omega \, .
\end{gather*}
\begin{lemma}
For the rescaled Schwarzschild metric (\ref{RescMet}), the values of $\Lambda$ and $\hat{\Phi}_{ab}$ are~:
\begin{eqnarray*}
\hat{\Lambda} &=& mR/2 \, ,\\
\hat{\Phi}_{ab} \d x^a \d x^b &=& \frac{1-3mR}{2} \left( R^2 F \d u^2 -2 \d u \d R + \d \omega^2  \right) \\
&=& \left( 1-3mR \right) \left( \frac{1}{2} \hat{g} + \d \omega^2 \right) \, .
\end{eqnarray*}
\end{lemma}
{\bf Proof.} The value of $\hat{\Lambda}$ was calculated in \cite{MaNi2009}. In order to evaluate $\hat{\Phi}_{ab}$, we determine $\hat{P}_{ab}$. First note that the Schwarzschild metric is Ricci flat, whence $\Phi_{ab}=P_{ab}=R_{ab}=0$. Hence,
\[ \hat{P}_{ab} = - \nabla_b \Upsilon_a  + \Upsilon_{AB'} \Upsilon_{BA'} \, .\]
Since $\Omega = R = 1/r$,
\[ \Upsilon_a \d x^a = -\frac{\d r}{r} \, .\]
We need to determine its spinor components. We do so in the dyad $\{ o^A , \iota^A \}$, denoting $x^0 = t$, $x^1 =r$, $x^2 = \theta$, $x^3 = \varphi$~:
\[ \Upsilon_{AA'} = \frac{-1}{r} g^1_{AA'} = \frac{-1}{r} g^{11} \varepsilon_{AB} \varepsilon_{A'B'} g_1^{BB'} = \frac{F}{r} \varepsilon_{AB} \varepsilon_{A'B'} g_1^{BB'} \]
and
\[ g_1^\mathbf{BB'} = \left( \begin{array}{cc} {n_1} & {-\bar{m}_1} \\ {-m_1} & {l_1} \end{array} \right) = \frac{1}{\sqrt{2F}} \left( \begin{array}{cc} {1} & {0} \\ {0} & {-1} \end{array} \right) \, . \]
It follows that
\[ \Upsilon_\mathbf{AA'} = -\frac{1}{r} \sqrt{\frac{F}{2}} \left( \begin{array}{cc} {1} & {0} \\ {0} & {-1} \end{array} \right) \, .\]
The non zero components of $\alpha_{ab} := \Upsilon_{AB'} \Upsilon_{BA'}$ are therefore
\begin{gather*}
\alpha_{00'00'} = \Upsilon_{00'} \Upsilon_{00'} = \alpha_{11'11'} = \Upsilon_{11'} \Upsilon_{11'} = \frac{F}{2r^2} \, , \\
\alpha_{01'10'} = \Upsilon_{00'} \Upsilon_{11'} = \alpha_{10'01'} = \Upsilon_{11'} \Upsilon_{00'} = -\frac{F}{2r^2}\, .
\end{gather*}
and the $2$-form $\alpha_{ab}$ reads~:
\begin{eqnarray*}
\Upsilon_{AB'} \Upsilon_{BA'}\d x^a \d x^b &=& \frac{F}{2r^2} \left( l_a l_b + n_a n_b - m_a \bar{m}_b - \bar{m}_a m_b \right) \d x^a \d x^b \\
&=& \frac{F}{2} \left( R^2 \hat{l}_a \hat{l}_b + R^{-2} \hat{n}_a \hat{n}_b - \hat{m}_a \bar{\hat{m}}_b - \bar{\hat{m}}_a \hat{m}_b \right) \d x^a \d x^b \\
&=& \frac{R^2F^2}{2} d u^2 -F \d u \d R + \frac{\d R^2}{R^2} - \frac{F}{2} \d \omega^2 \, .
\end{eqnarray*}
We now calculate $\nabla_b \Upsilon_a$~:
\[ \nabla_b \Upsilon_a \d x^a \d x^b = \nabla_b \left( -\frac{\d r}{r} \right) \d x^b = \frac{\d r^2}{r^2} + \frac{1}{r} \Gamma^1_\mathbf{ab} \d x^\mathbf{a} \d x^\mathbf{b} \, , \]
and among the Christoffel symbols
\[ \Gamma^1_\mathbf{ab} = \frac{1}{2} g^{1\mathbf{c}} \left( \frac{\partial g_\mathbf{ac}}{\partial x^\mathbf{b}} + \frac{\partial g_\mathbf{bc}}{\partial x^\mathbf{a}} - \frac{\partial g_\mathbf{ab}}{\partial x^\mathbf{c}} \right) =  -\frac{F}{2} \left( \frac{\partial g_{\mathbf{a}1}}{\partial x^\mathbf{b}} + \frac{\partial g_{\mathbf{b}1}}{\partial x^\mathbf{a}} - \frac{\partial g_\mathbf{ab}}{\partial r} \right) \, , \]
the non-zero ones are
\[ \Gamma^1_{00} = \frac{FF'}{2} \, ,~ \Gamma^1_{11} = -\frac{F'}{2F} \, ,~ \Gamma^1_{22} = -Fr \, ,~ \Gamma^1_{33} = -Fr \sin^2 \theta \, . \]
Hence
\begin{eqnarray*}
\nabla_b \Upsilon_a \d x^a \d x^b &=& \frac{\d r^2}{r^2} + \frac{FF'}{2r} \d t^2 -\frac{F'}{2rF} \d r^2 -F \d \omega^2 \\
&=& \frac{\d R^2}{R^2} +\frac{m}{r^3}\left( F\d t^2 - F^{-1} \d r^2 \right) - F \d \omega^2 \\
&=& FmR^3 \d u^2 -2mR \d u \d R + \frac{\d R^2}{R^2} -F \d \omega^2 \, .
\end{eqnarray*}
From this we can infer the value of $\hat{\Phi}_{ab}$~:
\begin{eqnarray*}
\hat{\Phi}_{ab} \d x^a \d x^b &=& \hat{P}_{ab} \d x^a \d x^b + \hat{\Lambda} \hat{g} \\
&=& \Upsilon_{AB'} \Upsilon_{BA'} - \nabla_b \Upsilon_a  + \frac{mR}{2} \hat{g} \\
&=&  \frac{1-3mR}{2} \left( R^2 F \d u^2 -2 \d u \d R + \d \omega^2  \right)  \, .
\end{eqnarray*}
This concludes the proof. \qed

\subsection{Energy estimate for the transverse derivative of a Dirac field}

First we set up a general formula for the commutation of a directional covariant derivative into the Dirac equation, then we apply it to the derivative along $\partial_R$.

Consider any vector field $V^a$ and commute $\hnabla_V$ into the Weyl equation~:
\begin{eqnarray*}
0= V^a \hnabla_a \hnabla^{BB'} \hat{\psi}_B &=& -V^a \hnabla_a \hat\varepsilon^{B'C'} \hnabla_{CC'} \hat{\psi}^C \\
&=& - \hat\varepsilon^{B'C'} \left( V^a \Delta_{ac} \hat{\psi}^C + \hnabla_{CC'} \left( V^a \hnabla_a \hat{\psi}^C \right) - \left( \hnabla_c V^a \right) \hnabla_a \hat{\psi}^C \right) \\
&=& - \hat\varepsilon^{B'C'} V^a \Delta_{ac} \hat{\psi}^C + \hnabla^{BB'} \left( \hnabla_V \hat{\psi}_B \right) - \left( \hnabla^b V^a \right) \hnabla_a \hat{\psi}_B
\end{eqnarray*}
where $\Delta_{ab} = \hnabla_a \hnabla_b - \hnabla_b \hnabla_a$. The first and third terms in the right-hand side can be calculated more explicitely~:
\[ V^a \hat{\Delta}_{ac} \hat{\psi}^C = V^a \left[ \hat{\varepsilon}_{A'C'} {\hat X}_{ACE}^{\Csp \Csp \Csp C} + \hat{\varepsilon}_{CA} \hat{\Phi}_{A'C'E}^{\CPsp \CPsp \Csp C} \right] \hat{\psi}^E \, . \]
The symmetries of the Riemann tensor imply that
\[ {\hat X}_{ACE}^{\Csp \Csp \Csp C} = 3 \hat{\Lambda} \hat{\varepsilon}_{AE} = \frac{3mR}{2} \hat{\varepsilon}_{AE} \, ;\]
whence
\[ - \varepsilon^{B'C'} V^a \hat{\varepsilon}_{A'C'} \hat{X}_{ACE}^{\Csp \Csp \Csp C} \hat{\psi}^E = \varepsilon^{B'C'} \frac{3mR}{2} V^A_{C'} \hat{\psi}_A = \frac{3mR}{2} V^{AB'} \hat{\psi}_A\, . \]
The term involving $\hat\Phi_{ab}$ can be written
\[ - \varepsilon^{B'C'} \hat{\varepsilon}_{AC} \hat{\Phi}_{A'C'E}^{\CPsp \CPsp \Csp C} \hat{\psi}^E = -V_{CA'} \hat{\Phi}^{EB'CA'} \hat{\psi}_E  = -V_a \hat{\Phi}^{BB'a} \hat{\psi}_B \, .\]
It follows that the equation satisfied by $\hnabla_V \hat{\psi}^B$ is
\begin{equation} \label{DirEqNablaVPsi}
\hnabla^{BB'} \left( \hnabla_V \hat{\psi}_B \right) = \left( \hnabla^b V^a \right) \hnabla_a \hat{\psi}_B  + V_a \hat{\Phi}^{ab} \hat{\psi}_B - \frac{3 mR}{2} V^b \hat{\Psi}_B  \, .
\end{equation}
If the vector field is $V^a \partial_a = \partial_R$, then
\[ \hnabla^\mathbf{a} V^\mathbf{b} = \hat{g}^{\ba \bd} \partial_\mathbf{d} V^\mathbf{b} + \hat{g}^{\ba \bd} \hat{\Gamma}^{\bb}_{\bd \bc} V^\bc = \hat{g}^{\ba \bd} \hat{\Gamma}^{\bb}_{\bd 1} \]
and the only non-zero coefficient is
\[ \hat{\Gamma}^1_{01} = R (1-3mR) \, .\]
Hence,
\[ \hnabla^a V^b \partial_a \partial_b = -R (1-3mR) \partial_R \otimes \partial_R  \]
and
\[ \left( \hnabla^b V^a \right) \hnabla_a \hat{\psi}_B = -R (1-3mR) V^b \hnabla_R \hat{\psi}_B \, . \]
The spinor $\hnabla_R \hpsi_A$ thus satifies
\begin{gather*} \label{DirEqNablaRPsi}
\hnabla^{AA'} \left( \hnabla_R \hat{\psi}_A \right) = \left[ \hnabla^{AA'} , \hnabla_R \right] \hat{\psi}_A = -R V^a \left( (1-3mR) \hnabla_R \hat{\psi}_A + \frac{3 m}{2} \hat{\Psi}_A \right) - \hat{\Phi}^{0a} \hat{\psi}_A \, , \\ \mbox{and } \hat{\Phi}^{0a} \partial_a = -\frac{1-3mR}2 \partial_R \, . \nonumber
\end{gather*}
The equation satisfied by higher order radial derivatives is obtained by means of the commutator expansion~:
\begin{eqnarray}
\hnabla^{AA'} \left( \hnabla_R^k \hat{\psi}_A \right) &=& \left[ \hnabla^{AA'} , \hnabla_R^k \right] \hat{\psi}_A \nonumber \\
&=& \sum_{p=0}^{k-1} \hnabla_R^{k-p-1} \left[  \hnabla^{AA'} , \hnabla_R \right] \hnabla_R^p \hat{\psi}_A \nonumber \\
&=& -\sum_{p=0}^{k-1} \hnabla_R^{k-p-1} \left( R V^a (1-3mR) \hnabla_R^{p+1} \hat{\psi}_A \right) \nonumber \\
&& + \sum_{p=0}^{k-1} \hnabla_R^{k-p-1} \left( \frac{1-6 mR}{2} V^a \hnabla_R^p\hat{\psi}_A \right)   \, . \label{DirEqHigherRDerivative}
\end{eqnarray}
It is useful to obtain the explicit expression of the action of $\hat{\Phi}^{0a}$ and $\partial_R$ on $\hat{\psi}_A$ by contraction. We simply need the Infeld-Van der Waerden symbol~:
\[ \hat{g}_1^\mathbf{AA'} = \left( \begin{array}{cc} {\hat{n}_1} & {-\bar{\hat{m}}_1} \\ {-\hat{m}_1} & {\hat{l}_1} \end{array} \right) = -\sqrt{\frac{2}{F}} \left( \begin{array}{cc} 1 & 0\\ 0 & 0 \end{array} \right) \, . \]
It follows
\begin{eqnarray*}
\left( \partial_R \right)^a \hat{\psi}_A &=& \hat{g}_1^{AA'} \hat{\psi}_A = - \sqrt{\frac2F} \hat{\Psi}_0 o^{A'}  \, , \\
\hat{\Phi}^{0a} \hat{\psi}_A &=& -\frac{1-3mR}2 \left( \partial_R \right)^a \hat{\psi}_A = \frac{1-3mR}{\sqrt{2F} }\hat{\Psi}_0 o^{A'} \, , \\
\end{eqnarray*}

The conservation law for $\hat{\nabla}_R \hat{\psi}_A$ is the following
\begin{eqnarray*}
\hnabla^{AA'} \left( ( \hnabla_R \hat{\psi}_A ) (\hnabla_R \bar{\hat{\psi}}_{A'} ) \right) &=& 2 \Re \left( -R (1-3mR) \left( \partial_R \right)^a (\hnabla_R \hat{\psi}_A ) \hnabla_R \bar{\hat{\psi}}_{A'} ) \right. \\
&& \left.  + \frac{1-6mR}{2} \left( \partial_R \right)^a \hat{\psi}_A \hnabla_R \bar{\hat{\psi}}_{A'} ) \right) \\
&=& \sqrt{\frac{2}{F}} \Re \left( 2R (1-3mR) \left\vert (\hnabla_R \hat\Psi )_0 \right\vert^2 - (1-6mR) \hat{\psi}_0 \overline{(\hnabla_R \hat{\Psi})_{0} } \right) \, .
\end{eqnarray*}
The $L^2$ norm of $\hat\psi_0$ on a $2$-sphere is controlled uniformally on $\Omega^+_{u_0}$ by that of $\eth' \hat\psi_0$. This in turn is controlled by the $L^2$ norms of $\hat\psi_1$ and $\partial_R \hat\psi_1$ using the second part of Dirac's equation. And since
\begin{eqnarray*}
(\hnabla_R \hat\Psi )_1 &=& (\hnabla_R \hat\Psi_A ) \hat\iota^A \\
&=& - \sqrt{\frac{2}{F}} \left( \hat{D} \hat{\psi}_A \right) \hat\iota^A \\
&=& - \sqrt{\frac{2}{F}} \left( \hat{D} \left( \hat{\psi}_1 \hat{o}_A - \hat{\psi}_0 \hat\iota_A \right) \right) \hat\iota^A \\
&=& - \sqrt{\frac{2}{F}} \left( \hat{\psi}_1 \hat\iota^A \hat{D} \hat{o}_A + \hat{D} \hat\psi_1 - \hat\psi_0 \hat\iota^A \hat{D} \hat\iota_A \right) \\
&=& - \sqrt{\frac{2}{F}} \left( \hat{\varepsilon} \hat{\psi}_1 + \hat{D} \hat\psi_1 - \hat{\pi} \hat\psi_0 \right) = \partial_R \hat\psi_1 - \frac{m}{2F} \hat\psi_1 \, ,
\end{eqnarray*}
we can estimate the $L^2$ norm of $\hat\psi_0$ on $S^2$ uniformly on $\Omega^+_{u_0}$ by the sum of the $L^2$ norms of $\hat\psi_1$ and $(\hnabla_R \hat\psi)_1$. Then we deal with the error term as we did in section \ref{Peeling}.
\begin{remark} \label{PurelyTransverse}
Note that the error term is much simpler than it was when we used partial derivatives. This is to be expected but the practical upshot here is that we do not need to control angular derivatives in order to get estimates on the derivative transverse to $\scri^+$. This remains true for higher orders. It can be useful to bear this in mind if one is interested in controlling transverse derivatives with low angular regularity. The same is very probably true for Maxwell.
\end{remark}

\end{document}